\documentclass[fleqn,usenatbib]{mnras}

\usepackage[T1]{fontenc}

\DeclareRobustCommand{\VAN}[3]{#2}
\let\VANthebibliography\thebibliography
\def\thebibliography{\DeclareRobustCommand{\VAN}[3]{##3}\VANthebibliography}

\usepackage{graphicx}	
\usepackage{amsmath}	
\usepackage{amssymb}	
\usepackage[export]{adjustbox}
\usepackage{xcolor}
\usepackage{caption}
\usepackage{siunitx}
\usepackage{subfigure}

\newcommand{\de}{{\rm d}}
\newcommand{\phz}{photo--$z$~}
\newcommand{\phzs}{photo--$z$'s~}

\usepackage{eso-pic}

\AddToShipoutPictureBG*{%
 \AtPageUpperLeft{%
  \hspace{0.65\paperwidth}%
  \raisebox{-4.5\baselineskip}{%
}}}%

\title[Standard siren measurement of $H_0$ using O1-O4a GW events with Legacy]{A dark standard siren measurement of the Hubble constant following LIGO/Virgo/KAGRA O4a and previous runs}

\author[ C.R. Bom \& V. Alfradique et al.]{
C. R. Bom$^{1,2}$\thanks{E-mail: debom@cbpf.br}
V. Alfradique$^{1}$,
A. Palmese$^{3}$,
G. Teixeira$^{1}$,
L. Santana-Silva$^{1}$,
A. Santos$^{1}$,
\newauthor
P. Darc$^{1}$
\\
\\
$^{1}$Centro Brasileiro de Pesquisas F\'isicas, Rua Dr. Xavier Sigaud 150, 22290-180 Rio de Janeiro, RJ, Brazil \\
$^{2}$Centro Federal de Educa\c{c}\~{a}o Tecnol\'{o}gica Celso Suckow da Fonseca,  Rodovia M\'{a}rcio Covas, lote J2, quadra J - Itagua\'{i} (Brazil)\\
$^{3}$McWilliams Center for Cosmology, Carnegie Mellon University, 5000 Forbes Ave, Pittsburgh, PA 15213, USA 
}
\date{Accepted XXX. Received YYY; in original form ZZZ}

\pubyear{2024}

\begin{document}
\label{firstpage}
\pagerange{\pageref{firstpage}--\pageref{lastpage}}
\maketitle

\begin{abstract}
We present a new constraint on the Hubble constant ($H_0$) from the standard dark siren method using a sample of  $5$ well-covered gravitational waves (GW) alerts reported during the first part of the fourth LIGO/Virgo/KAGRA observing run and with $3$ updated standard dark sirens from third observation run in combination the previous constraints from the first three runs. Our methodology relies on the galaxy catalogue method alone. We use a Deep Learning method to derive the full probability density estimation of photometric redshifts using the Legacy Survey catalogues. We add the constraints from well localized Binary Black Hole mergers to the sample of standard dark sirens analysed in our previous work. We combine the $H_0$ posterior for $5$ new standard sirens with other $10$ previous events (using the most recent available data for the 5 novel events and updated 3 previous posteriors from O3), finding $H_0 =  70.4^{+13.6}_{-11.7}~{\rm km~s^{-1}~Mpc^{-1}}$ (68\% Confidence interval) with the catalogue method only.  This result represents an improvement of $\sim 23\%$ comparing the new $15$ dark siren constrain with the previous $10$ dark siren constraint and a reduction in uncertainty of $\sim 40\%$ from the combination of $15$ dark and bright sirens compared with the GW170817 bright siren alone.
The combination of dark and bright siren GW170817 with recent jet constraints yields $H_0$ of $68.0^{+4.4}_{-3.8}~{\rm km~s^{-1}~Mpc^{-1}}$,  a $\sim 6\%$ precision from Standard Sirens, reducing the previous constraint uncertainty by $\sim 10\%$ .
\end{abstract}





\section{Introduction}
\label{sec:intro}
The current $4-6\sigma$ tension in the Hubble constant \citep{Riess_2019,planck18,Freedman_2019,riess2021} arises from the significant discrepancy between different cosmological probes, in particular from the Cosmic Microwave Background \cite[CMB,][]{planck18}, and those using Supernovae (SN) and Cepheids for the local distance ladder \citep{riess2021}. New independent measurements of the Hubble constant have the potential to shed light on this discrepancy (e.g., \cite{verde,Dainotti,abdalla2022cosmology}), and, depending on their precision, could arbitrate the tension. Among novel probes, the standard sirens (see \cite{schutz}) methodology employs gravitational wave occurrences to obtain luminosity distances. This information is used to infer cosmological parameters, most notably the Hubble constant  $H_0$, upon integration with redshift data derived from host galaxies. This emergent probe could play an important role as it is independent of the cosmic distance ladder \citep{chen17,gray2019cosmological,bom23bbh,10.1093/mnras/stae1360, 2021PhRvD.103d3520M,2022MNRAS.511.2782C}.

Standard sirens are categorized as "bright" in instances where an electromagnetic counterpart is definitively identified alongside a singular host galaxy, and as "dark" or "statistical" in the absence of such counterparts. Bright sirens, exemplified by GW170817 \citep{ligobns}, yield measurements of high precision. However, the requirement of host identification poses a series of challenges due to the wide search volume and the cadence requirements \citep{bom2024strategy,andreoni2022strategy}. Furthermore, kilonovae have been the only widely confirmed electromagnetic sources detectable from gravitational waves, although black hole mergers are proposed to produce electromagnetic counterparts in certain circumstances with a few identified candidates \citep{grahamLightDarkSearching2023,2024arXiv240710698C,2024arXiv240709945R,2024MNRAS.527.6076R}. Therefore, the dark siren method can be applied to a larger number of events, including GW170817 \citep{fishbach}, the binary black hole mergers GW170814 \citep{darksiren1} and GW190814 \citep{palmese20_sts}, and several events from the first three LIGO/Virgo observing runs \citep{Abbott_2023}. Dark standard sirens assuming the catalog method \citep{Hitchhiker} rely on the position and redshift of potential host galaxies, leading to less precise results than bright sirens on a single-event basis. However, combining dark and bright sirens can enhance constraints on cosmological parameters, leveraging the abundance of events without counterparts.

The fourth LIGO/Virgo/KAGRA (LVK) Observing run (O4) began on May 24, 2023, and is scheduled to span 20 months, including two months allocated for commissioning breaks dedicated to maintenance.
The initial segment of O4 referred to as O4a, finished on January 16, 2024. During O4a there have been $82$ GW alerts reported, of which $80$ were classified as binary black hole candidates. It is important to note that the Virgo detector was not operational during the O4a run, therefore the typical sky localizations were worse than would have been expected if it had joined the run. Meanwhile, KAGRA has participated in the run for a limited duration, albeit at a notably reduced binary neutron star (BNS) inspiral range compared to the LIGO detectors.
In this work, we use a set of well-localized and confident candidate events from O4a covered by public photometric catalogs and imaging data, mainly from the Legacy Survey \citep[LS; ][]{zhou2020clustering,dey2019} and DECam Local Volume Exploration Survey~\citep[DELVE;][]{delvedr1,delvedr2}.
The photometric redshifts were computed using the same deep learning technique from \cite{alfradique2024}. We adding 5 new sirens to the sample of dark sirens from \citep{palmese20_sts,palmese2023,alfradique2024}, increasing the total to $15$ dark sirens. The dark siren catalog method used in this work to constrain $H_0$ is described in the aforementioned papers, and we also refer to the section \ref{sec:method} for a description of our methodology.

\section{Data}\label{data}

\begin{figure*}
\centering
\includegraphics[width=1\linewidth]{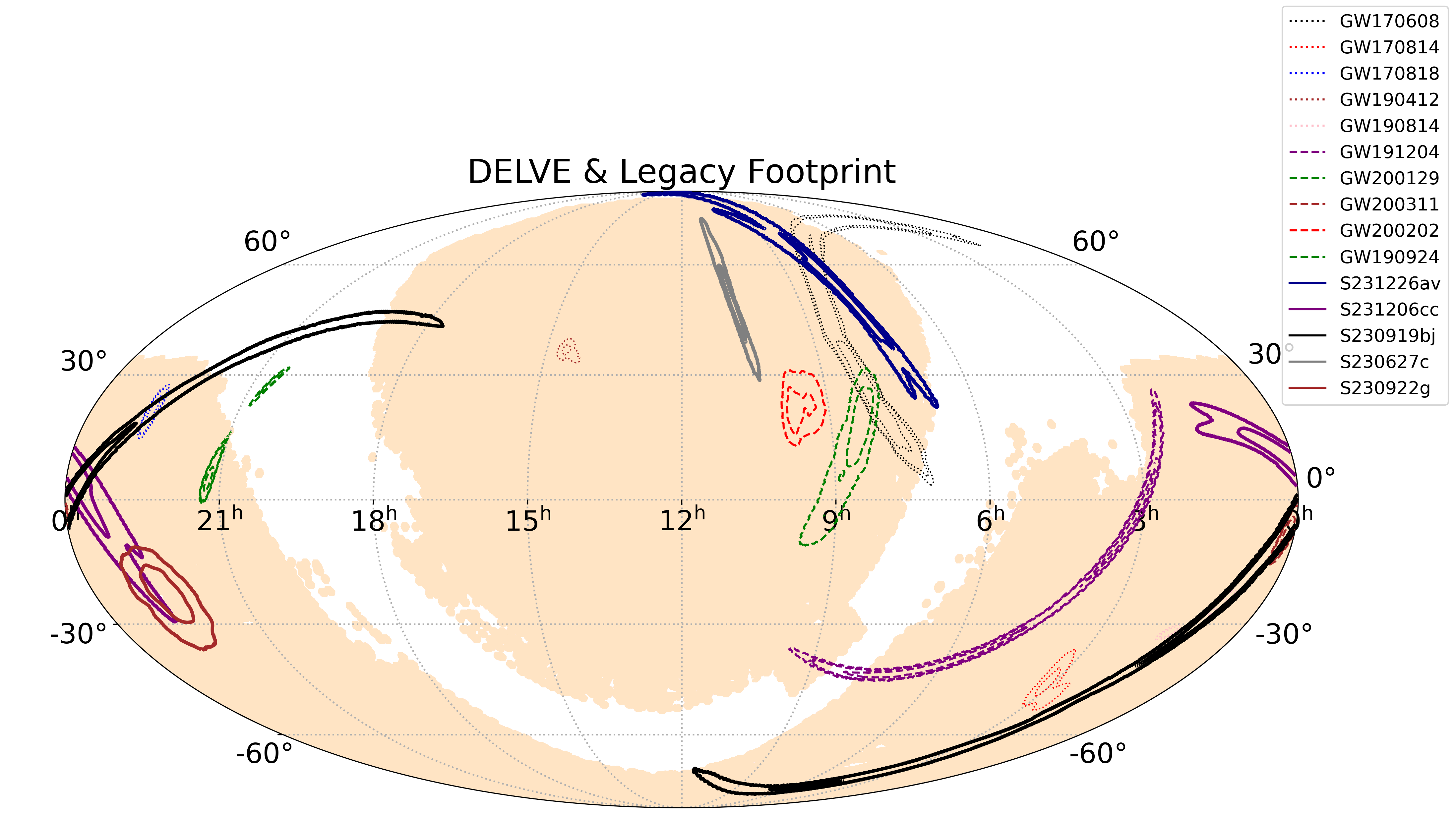}
\caption{The LIGO/Virgo/KAGRA dark sirens analysed in this paper. The contours depict the 90$\%$ CI localization from the sky maps. Dotted contour lines indicate events scrutinized by \citet{palmese2023}. Dashed lines represent events examined in \citet{alfradique2024}, which underwent reprocessing. Solid lines denote newly incorporated events in the analysis. The light orange shaded areas represent regions covered by the DELVE and DESI Legacy Survey catalogues.}
\label{fig:delvelegacy_footprint}
\end{figure*}

\subsection{The LIGO/Virgo GW data}\label{sec:GWdata}

In this study, we expand the analysis upon the previous works with 8 events \citep[][thereafter P23]{palmese2023} and 10 events \citep[][thereafter A24]{alfradique2024} from LVK run O1-O3, using the catalogue method. We employ comparable selection criteria from prior studies (see P23 and A24) and, as such, investigate five new sirens identified in O4 with more than 70$\%$ of their probability coverage falling within the Legacy or DELVE surveys, and with luminosity distances less than $d_L<1500\, \rm{Mpc}$. 
We restricted in distance based on the previous works, as the catalogs turn more incomplete and also the dependency on $\Omega_m$ becomes more relevant. The positional data for these events comes from the maps publicly provided by the LVK collaboration. We present the 90\% Credible Interval (CI) sky region of all events used in this work in figure~\ref{fig:delvelegacy_footprint}. These maps of right ascension (RA), declination (dec), and distance probability, are represented using HEALPIX pixelation. Within this framework, the probability distribution along each line of sight from the maps is assumed to follow a Gaussian distribution. We draw attention to the fact that the criteria adopted here are chosen with the intention of selecting those superevents with greater constraint capability, but there is no impediment for others to be added to the sample, as long as the selection function (see its definition in section \ref{sec:method}), defined in the $H_0$ posterior, correctly describes the cuts adopted.

The five novel siren events we selected are S231226av \citep{GCN_S231226av}, S231206cc \citep{S231206cc}, S230922g \citep{S230922g}, S230919bj \citep{S230919bj} and S230627c \citep{S230627c}. They are all classified as Binary Black Hole mergers (BBH) with probability $> 99\%$, except for S230627c, which has a classification of 49\% for neutron star-black hole, 48\% for BBH, and 3\% of being noise. In particular, S231226av is among the lowest False-Alarm-Rate (FAR) events, while S230919bj and S230627c are in the top $20\%$ percentile of low FAR in O4. The novel events $90\%$ volume are comparable to the previous sirens studied with $90\%$ volume $\sim 10^{-3}$ Gpc$^3$. We use the latest public skymaps (see table \ref{tab:events} for details) produced by the python code Bilby \citep{bilby}. We use the last skymaps from the GW Transient Catalogue (GWTC \footnote{\url{https://gwosc.org/eventapi/html/GWTC/}}, \cite{gwtc3}) for three events from P23, which used the maps from the GW alerts S191204r, S200129m, and S200311bg. The updated events in this work are named GW191204\_171526, GW200129\_065448 and GW200311\_115853 respectively. For the remaining events, we use the last publicly available skymaps on the GraceDB event page \footnote{\url{https://gracedb.ligo.org/}}.  Furthermore, to perform an additional check on the quality of the selected GW alerts from O4 in this work, we reproduced the same cuts used on the O3 GW alerts. These cuts excluded retracted events after a few hours or days, using the same level of FAR considered, distance and area.

\begin{table*}
\centering
\begin{tabular}{cccccc}
Event & $d_L$ [Mpc] & A [deg$^2$] & V [Gpc$^3$] & FAR  \\
\hline \hline
GW170608$^1$ & $320^{+120}_{-110}$ & 392 & $3 \times 10^{-3}$ &$<1$ per $ 10^5$ yr \\
GW170814$^2$ & $540^{+130}_{-210}$ & 62 & $2 \times 10^{-3}$ &$<1$ per $ 10^4-10^7$ yr \\ 
GW170818$^1$ & $1060^{+420}_{-380}$ & 39 & $7 \times 10^{-3}$ &$<1$ per $ 10^5$ yr \\
GW190412$^3$ & $740^{+120}_{-130}$ & 12 & $4 \times 10^{-4}$  &$<1$ per $ 10^3-10^5$ yr \\
GW190814$^4$ & $241^{+26}_{-26}$ & 19 & $3 \times 10^{-5}$ &$<1$ per $ 10^4-10^7$ yr \\
GW190924\_021846$^3$ & $564^{+145}_{-145}$ & 348 & $1 \times 10^{-2}$ & $<1$ per $ 10^{5}$ yr \\
GW191204\_171526$^5$ & $624^{+123}_{-123}$ & 256 & $8 \times 10^{-3}$ &$<1$ per $ 10^{5}$ yr \\
GW200129\_065448$^5$ & $929^{+179}_{-179}$ & 31 & $3 \times 10^{-3}$ & $<1$ per $ 10^{5}$ yr  \\
GW200311\_115853$^5$ & $1154^{+206}_{-206}$ & 35 & $6 \times 10^{-3}$ & $<1$ per $ 10^{5}$ yr  \\
GW200202\_154313$^8$ & $409^{+95}_{-95}$ & 167 & $2 \times 10^{-3}$ & $<1$ per $ 10^{5}$ yr \\
\hline \hline 
S231226av$^{6}$ & $1218^{+171}_{-171}$ & 199 & $3 \times 10^{-2}$ & $<1$ per $ 10^{42}$ yr \\  
S231206cc$^{7}$ & $1467^{+264}_{-264}$ & 342 & $9 \times 10^{-2}$ & $<1$ per $ 10^{27}$ yr \\
S230922g$^{8}$ & $1491^{+443}_{-443}$ & 324 & $1 \times 10^{-1}$ & $<1$ per $10^{16}$ yr \\
S230919bj$^{9}$ & $1491^{+402}_{-402}$ & 708 & $2 \times 10^{-1}$ & $<1$ per $ 10^{2}$ yr \\
S230627c$^{10}$ & $291^{+64}_{-64}$ & 82 & $4 \times 10^{-4}$ &  $<1$ per $ 10^{2}$ yr \\
\hline \hline
\end{tabular}
\caption{Luminosity distance, 90\% CI area and volume, and FAR of gravitational wave events and candidates used in this analysis. We also report the reference paper or GCN that reports to the sky map used for each event. Where a range of FAR is provided, this is because multiple FAR estimates are available from multiple search algorithms. The FAR reported for the candidates is different from the confirmed events as it is estimated from the online analysis. These candidates have all recently been confirmed as gravitational wave events in \citet{gwtc3}. References for each event: (1) \citet{gwtc1}, (2) \citet{gw170814}, (3) \citet{gwtc2}, (4) \citet{190814_paper}, (5) \citet{gwtc3_published}, (6) \citet{GCN_S231226av}, (7) \citet{GCN_S231206cc}, (8) \citet{GCN_S230922g} (9) \citet{GCN_S230919bj}, (10) \citet{GCN_S230627c}
     }
\label{tab:events}
\end{table*}

\subsection{Optical Survey Data}\label{sec:desi}

We made use of publicly available catalogues from the DESI Legacy Survey~\citep[][]{dey2019} and DELVE survey \citep{delvedr1,delvedr2}. The combined footprint is presented in figure \ref{fig:delvelegacy_footprint}. We use this data to obtain precise photometric redshifts using the same technique and Deep Learning method described in A24. We use a Mixture Density Network \citep[MDN, ][]{Bishop_1994} to derive the full probability density functions (PDFs) for each galaxy in the survey catalogue. A more comprehensive description of the model and photometric redshift quality assessment in both surveys and a comparison between the MDN method and the public data for the Legacy Survey is presented in the next subsection \ref{sec:a1_photoz}. Our final constraints use the Legacy Survey PDF instead of Gaussian approximations from the Legacy Survey photo-$z$ catalogues \citep{zhou2020clustering}. The Legacy Survey results slightly outperformed the DELVE for the same model. We use DELVE-based photometric redshift as a validation and we found a small impact for the final constraints of $< 0.5\, \rm{km/s/Mpc}$.

In the GW sample processed in this work, only the S230919bj event has significant 90\% region uncovered by the photo-$z$ catalogues from both Legacy galaxy survey and DELVE. To compensate for the insufficient coverage, we adopted the same procedure from \cite{palmese2023}. Our strategy involves distributing simulated galaxies in regions lacking data. To ensure that marginalization includes all possible host galaxies and leave our Hubble constant measurement free of underestimated uncertainty, the injected galaxies follow our prior distribution as given by the training sample. The photo--$z$ distribution of these fake galaxies was sampled using the Monte Carlo technique. We assume a uniform spatial distribution, and the number density is given by the value of the Legacy Survey galaxy catalogue. For the photo--$z$ precision, we first found the relation between the photo-$z$ error and photo-$z$ by computing the mean and standard deviation value of the photo-$z$ error in photo-$z$ bins of size 0.05 from our training sample, after that, we sampled from a Gaussian function for each photo-$z$ bins. This same procedure was adopted for the apparent magnitude distribution of these fake galaxies, since it is intrinsically associated with the redshift value. Here, we also assume that all the fake photometric redshift PDFs follow a Gaussian function; we make a Gaussian sampling around the assigned true value with the standard deviation given by their respective photo-$z$ error. In section \ref{results} we present the impact of these fakes galaxies and found it to be minor.

\subsection{Galaxy Catalogues and Photometric Redshifts}
\label{sec:a1_photoz}

In this work, we use catalogue data from both DEcam Local Volume Exploration survey \citep{delvedr1,delvedr2} and the Legacy Survey \citep{dey2019} as the main data used to select the galaxies within the LIGO/Virgo localization for each event. It is worth mentioning that the Legacy Survey catalogues incorporated publicly available DECam data, including DELVE footprint for the LS DR10.
The DELVE survey observed a large fraction of the Southern sky covering an area of 21,000 deg$^{2}$, of which 17,000 deg$^{2}$ were homogeneously observed in the four broad bands ({\it g}, {\it r}, {\it i}, and  {\it z}) and photometric depth up to 24.3, 23.9, 23.5, and 22.8 for {\it g},{\it r}, {\it i}, and {\it z}, respectively for 5$\sigma$ point source detections. The DELVE catalogues comprise $2.5$ billion sources, in which $618$ million have data in all the four band. The Legacy survey explored a significant portion of the sky ($\sim 33,500$ sq deg) in $grz$ bands, reaching depths of 24.0, 23.4, and 22.5 in $grz$ for 5$\sigma$ detection.
The sky coverage of DELVE and Legacy surveys catalogues are shown in figure \ref{fig:delvelegacy_footprint}, together with the 50$\%$ and 90$\%$ CI of the GW events studied in previous works and the five new events explored in this study.

Photometric redshifts used in this work were determined using the same deep learning method presented in \cite{alfradique2024}. 
The fundamental component of the Deep Learning model comprises a neural network that analyses tabular data, using a Legendre Memory Unit \citep{voelker2019lmu} with a MDN. Unlike traditional neural networks that provide single value estimations, MDNs outputs conditional probability densities through a linear combination of individual probability distributions (components), chosen to be Gaussian distributions in our case. This approach enables a more comprehensive characterization of predictions and errors assessment. 

The neural network output is a linear combination of $C$ Gaussian kernels ($\mathcal{N}(\mu_i, \sigma_i)$, where $\{\mu_i\}$ is the mean and $\{\sigma_i\}$ is the standard deviation) weighted by mixture coefficients $\{\alpha_i\}$. We impose, for the mixture coefficients, that  $\sum^C_{i=1}\alpha_i=1$ and $0<\alpha_i<1 $. Therefore, the PDF can be written as

\begin{equation}
\label{eq:mdn_final_pdf}
    \rm{PDF}(z) = \sum^C_{i=1} \alpha_i \mathcal{N}(\mu_i, \sigma_i).
\end{equation}

One PDF is assigned to each galaxy, i.e., the MDN model predicts different sets of $\{\mu_i\}$, $\{\sigma_i\}$, and $\{\alpha_i\}$ for different photometric inputs.
This method was implemented to estimate \phzs PDFs for the DELVE DR2 and LS DR10 catalogues. The architecture used for the DELVE \phz is detailed in \cite{teixeira2024}. The LS DR10\footnote{ We used the latest available version of DR10, namely $10.1$.} \phz were estimated using a similar architecture, with minor adjustments after a fine tunning considering the point statistics metrics such as reducing the width of the dense layers and reducing the number of components $C$ in the mixture, from $C=20$ for DELVE to $C=6$ for Legacy. Input features for this process include \textit{griz} magnitudes and colour indices (\textit{g-r}, \textit{g-i}, \textit{g-z}, \textit{r-i}, \textit{r-z}, and \textit{i-z}). The spectroscopic data used as training set for DELVE catalogues is the same as the one presented in \cite{teixeira2024}, which was created from the crossmatch between DELVE catalogue and several spectroscopic data available in different large sky surveys. For the Legacy survey, we use the same training sample from the public photometric redshift catalogue \citep{zhou2020clustering}.

The LS public data releases do not directly provides the apparent magnitude in photometric band. Therefore, we initially used the linear fluxes (columns \texttt{FLUX\_\{G,R,I,Z\}}) to compute the magnitudes $m$ by employing the conversion $m=22.5-2.5\log_{10}(f)$, and derived the magnitude errors also from the inverse variances of the fluxes\footnote{See the description of the \texttt{FLUX\_IVAR\_\{G,R,I,Z\}} flag in \href{https://www.legacysurvey.org/dr9/description/\#photometry}{https://www.legacysurvey.org/dr9/description/photometry}.}. To mitigate star contamination, we adopted the same approach outlined by \cite{palmese2023}, applying colour cuts based on GAIA data, such as removing all known stars. Additionally, all the magnitudes were corrected for Milky Way extinction. Finally, we restrict our GW analysis to \textit{r} band magnitudes to be lower than 21. 

We conducted a series of selection cuts based on the photometry quality and properties to ensure the utilization of galaxies with the best possible detections. We excluded all objects with unphysical colours, retaining only those that satisfy the conditions:
$$-1> g-r,\; r-i,\/ i-z <4$$

For the spectroscopic sample (refer to \cite{teixeira2024} for a detailed list of the spectroscopic catalogues also used in this work), we restricted our objects to $0.01<z_{\rm{spec}}<1.5$. After applying these cuts, our spectroscopic sample contains $\sim 2.2$M galaxies.  We selected the training sample in order to have a uniform $z_{\rm{spec}}$ distribution of 0.01 to 1.0, and comprise all objects available with $z_{\rm{spec}}>1$ (which are few in number), resulting in 550k (580k for DR9) galaxies for training the model. The same approach of uniform training was also used in Legacy Survey publicly available photo-$z$ \citep{Zou_2019} to avoid possible systematic bias towards oversampled regions.

\begin{figure*}
\centering
\includegraphics[width=0.32\linewidth]{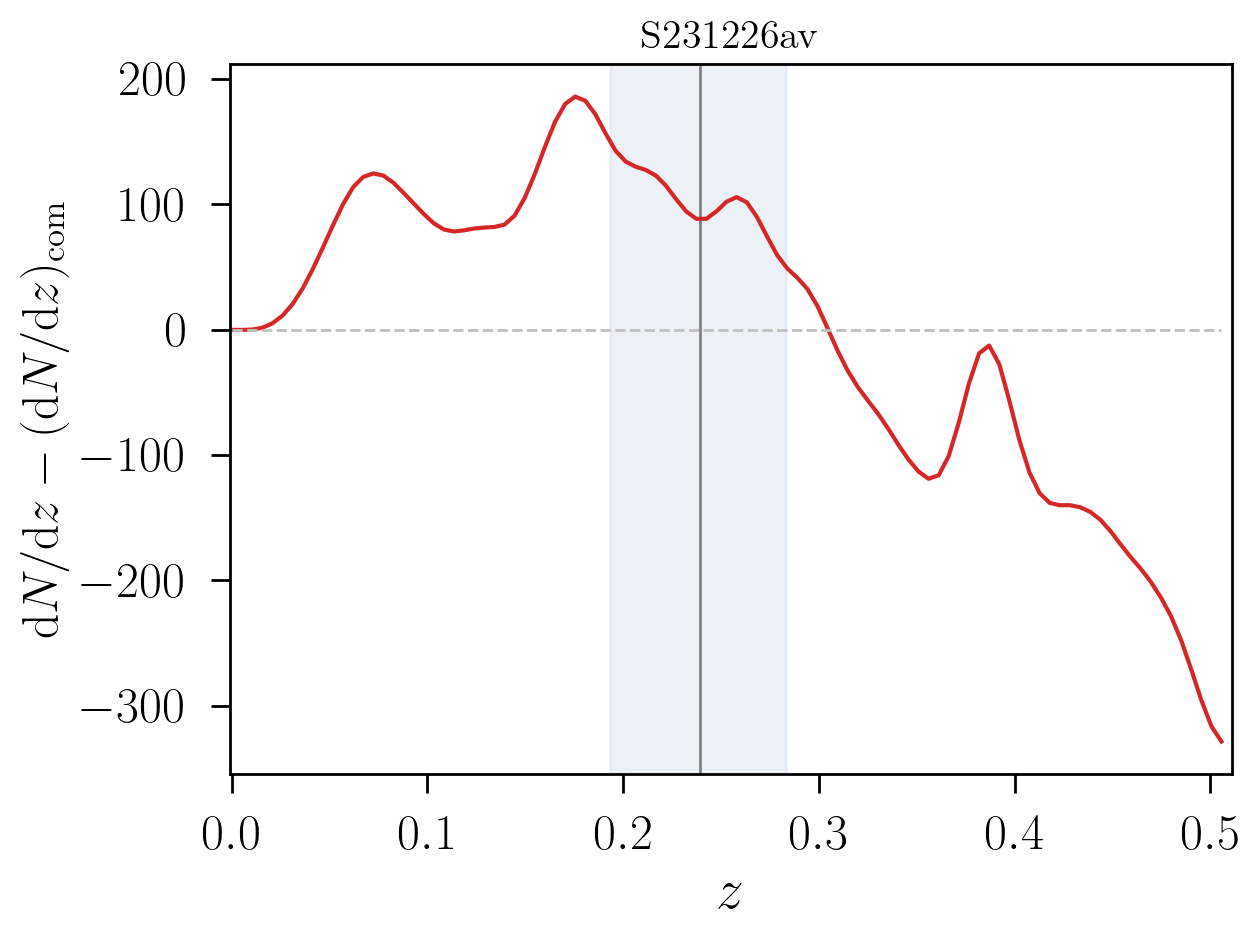}
\includegraphics[width=0.32\linewidth]{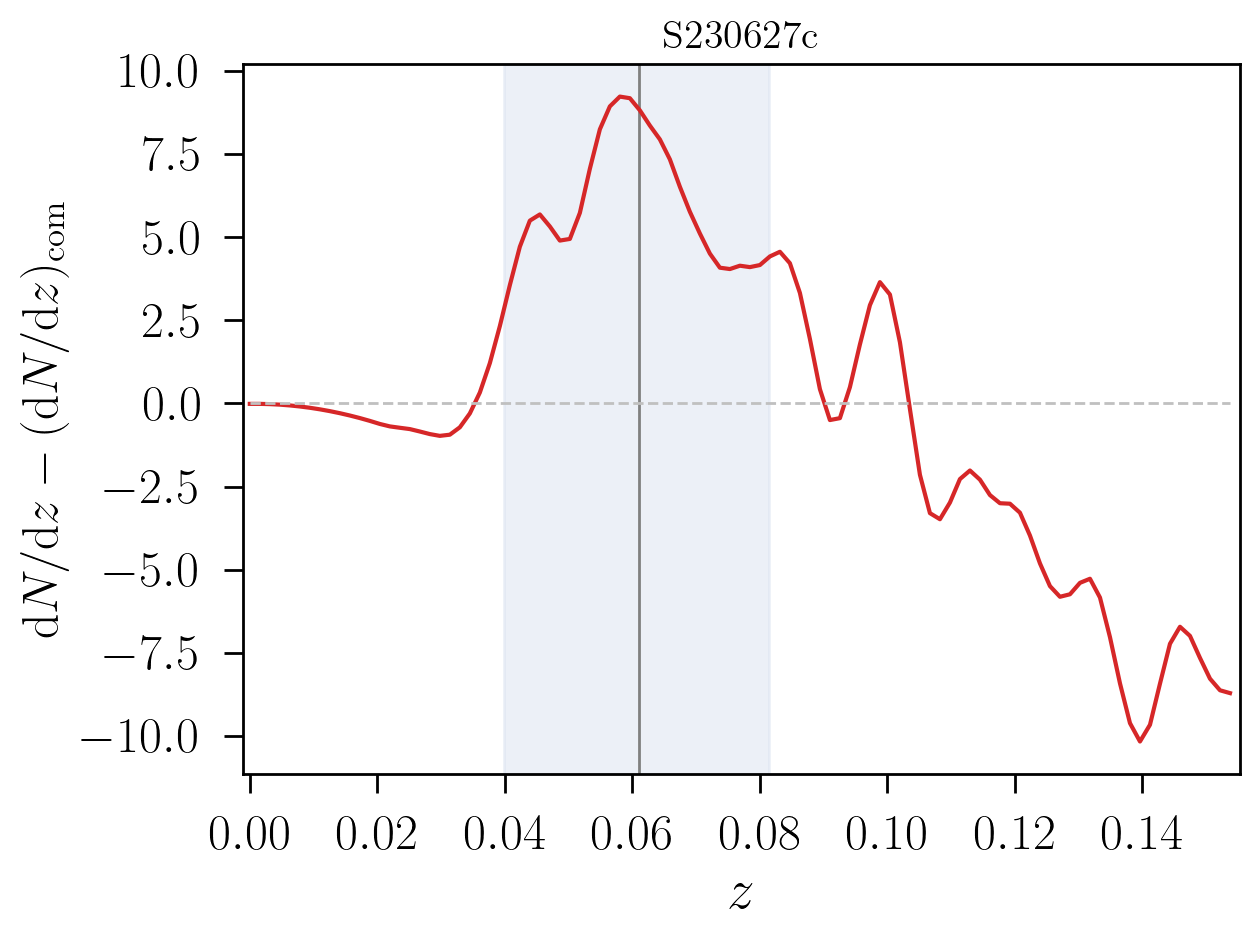}
\includegraphics[width=0.32\linewidth]{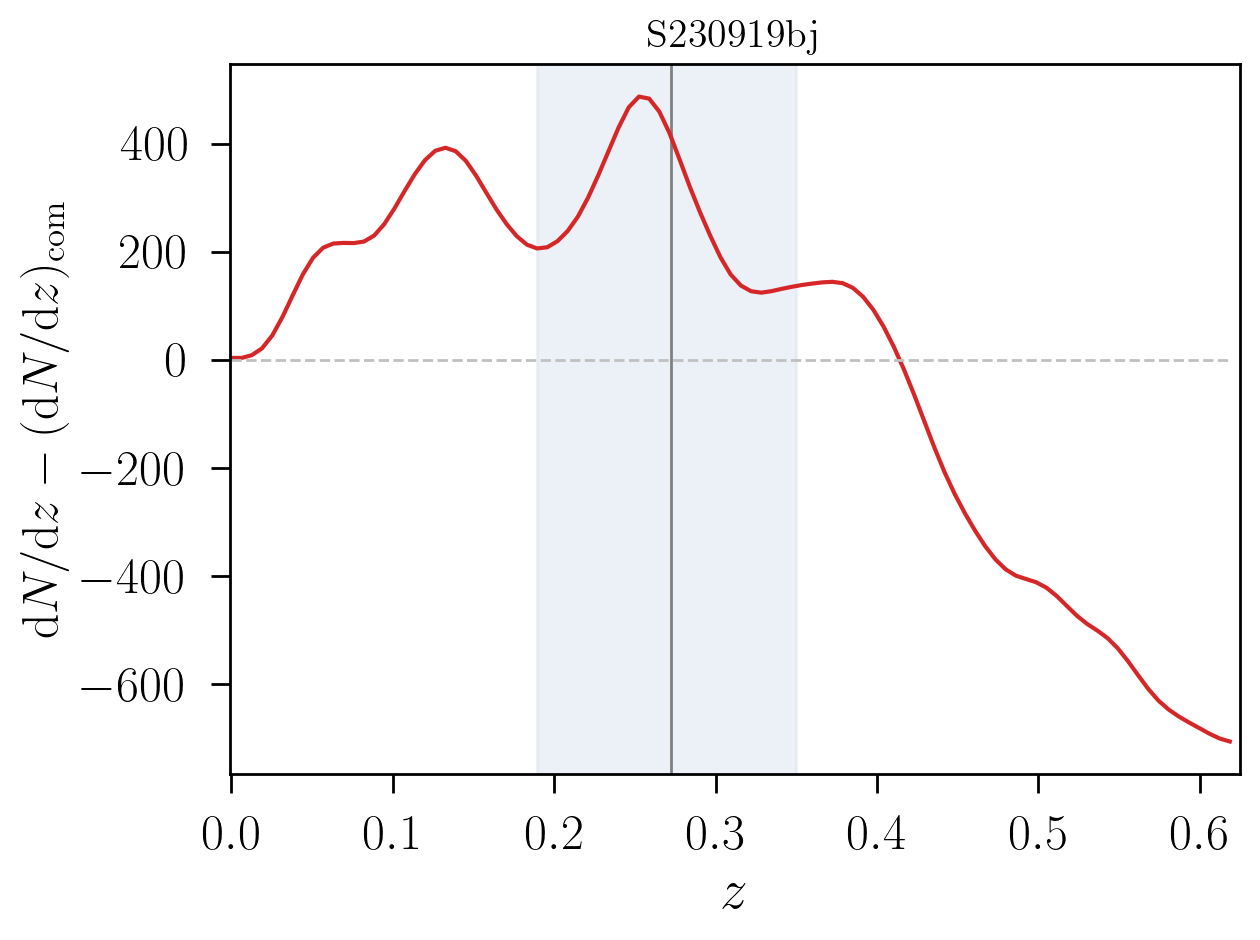}\\
\includegraphics[width=0.32\linewidth]{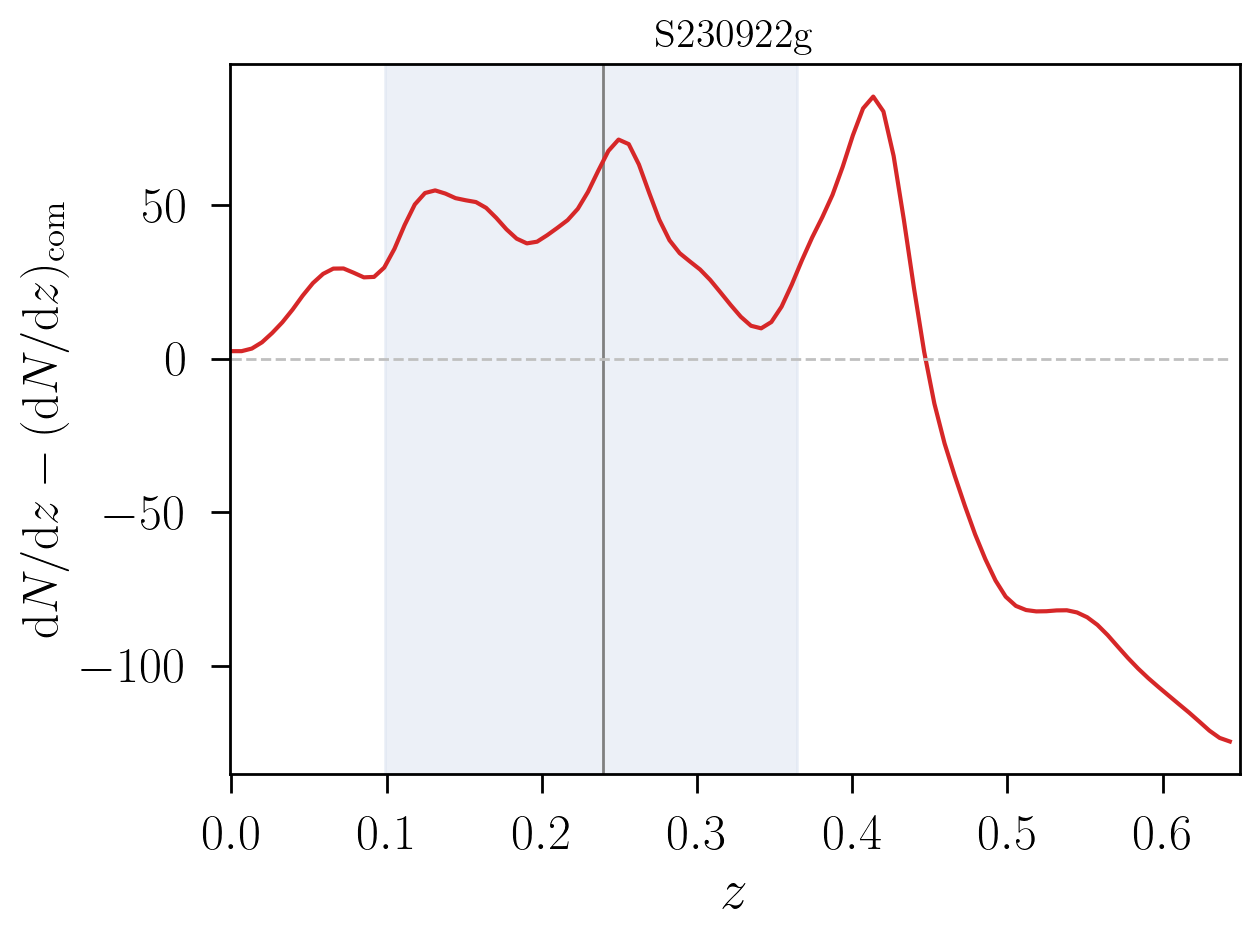}
\includegraphics[width=0.32\linewidth]{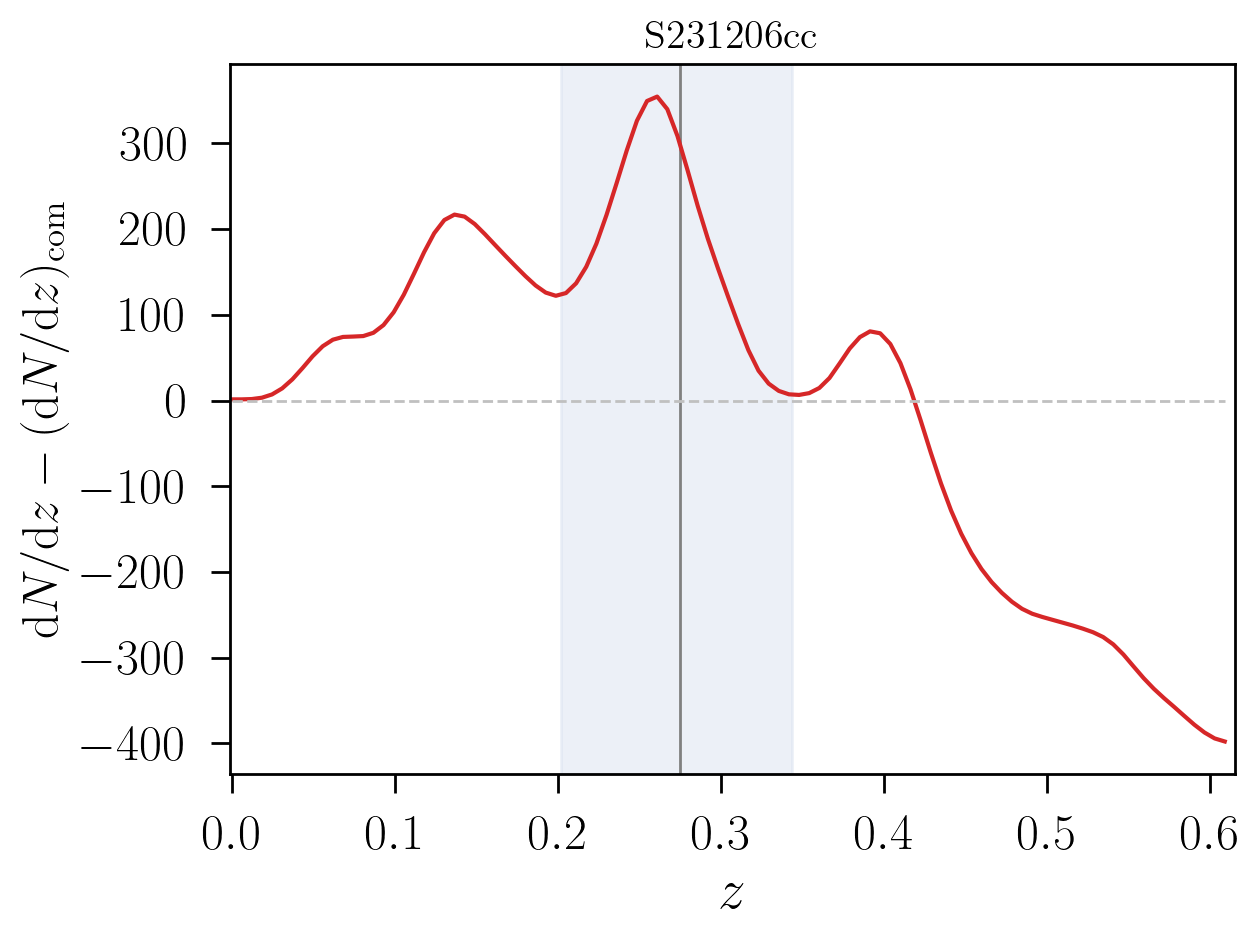}
\includegraphics[width=0.32\linewidth]{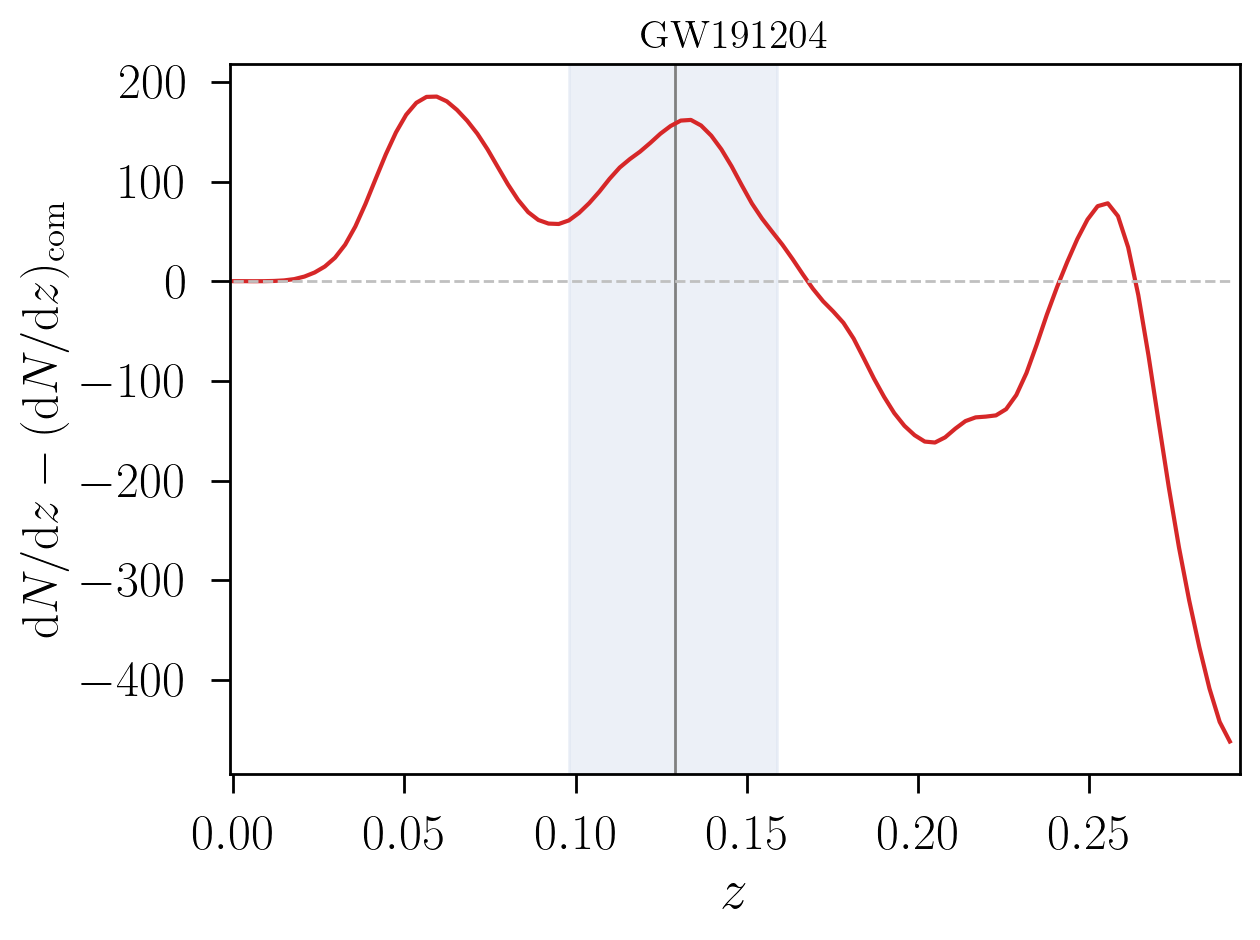}\\
\includegraphics[width=0.32\linewidth]{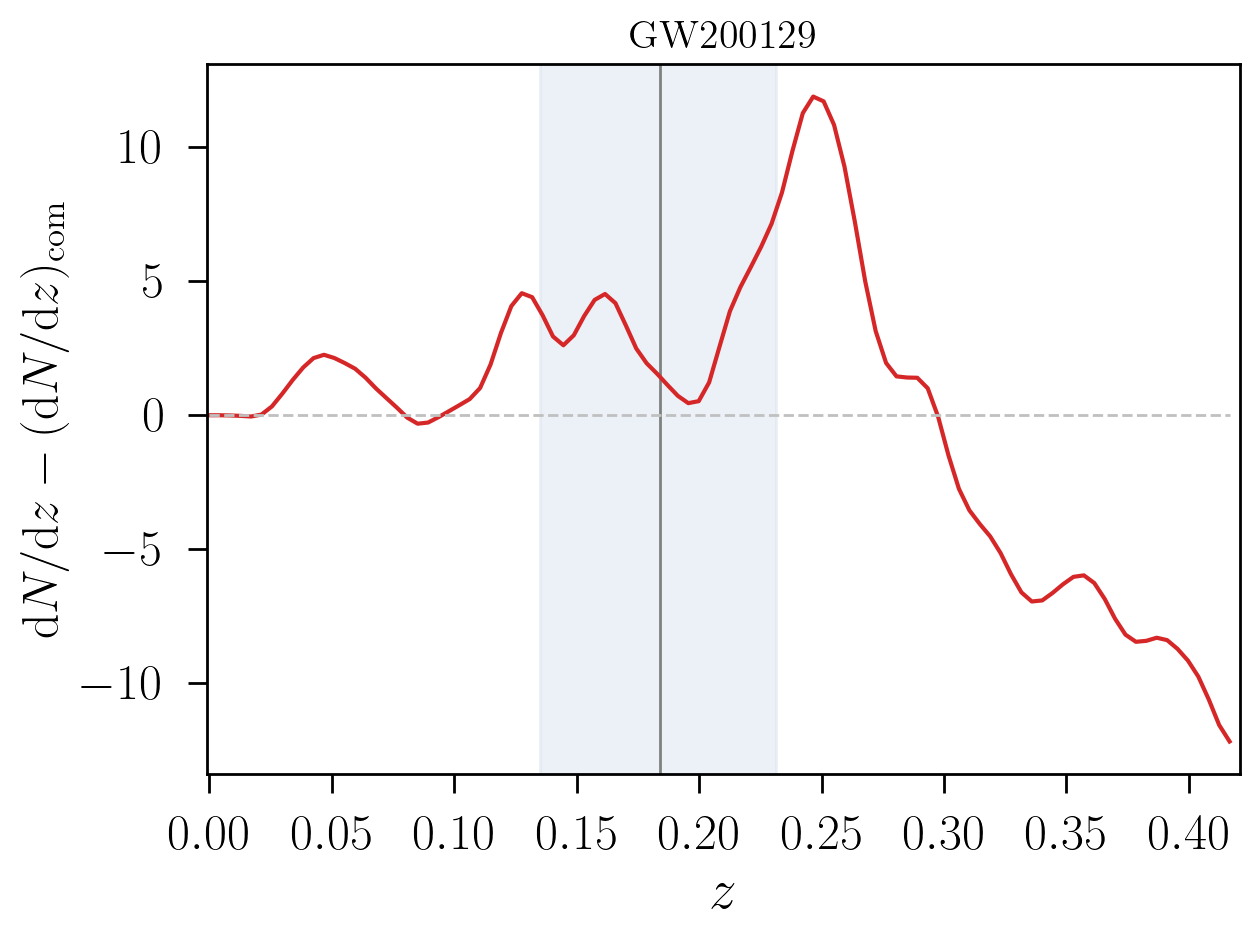}
\includegraphics[width=0.32\linewidth]{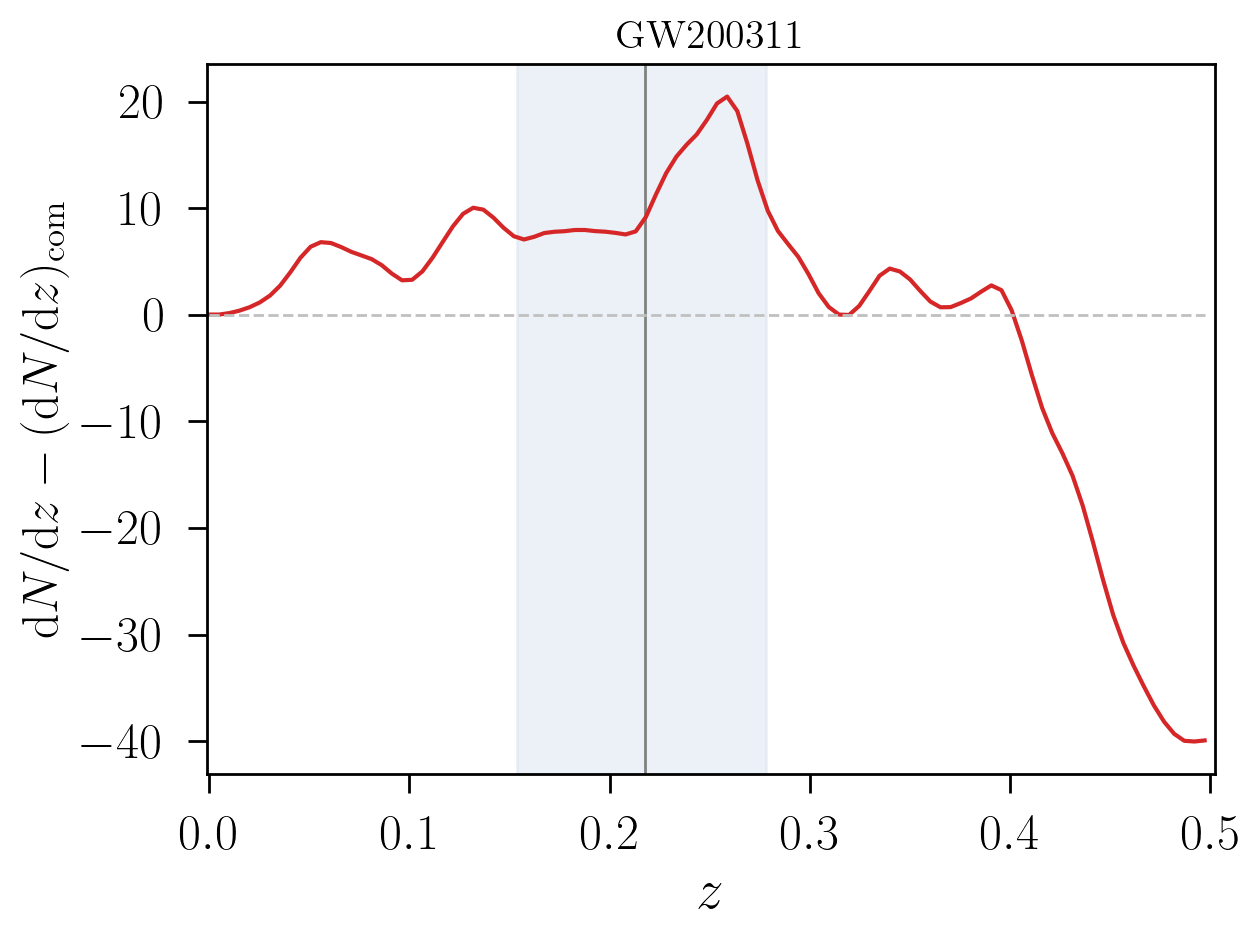}\\
\caption{Redshift distribution of galaxies in the 90\% CI area of the five new dark siren events from O4a (S231226av, S230627c, S230919bj, S230922g, and S231206cc), and the three superevents discussed in \citet{palmese2023} that have now been confirmed as gravitational waves (GW191204, GW200129 and GW200311). To highlight the presence of overdensities and underdensities along the line of sight, the redshift distribution was subtracted with a uniform number density. The grey vertical lines represent the luminosity distance of each GW event marginalized over the entire sky, assuming an $H_0$ of 70 km/s/Mpc, and the shaded regions are the $1\sigma$ uncertainties considering the same $H_0$. These regions are only showed for reference.}
\label{fig:dndz}
\end{figure*}

We generate PDFs for galaxies in the test sample to validate the model by checking probabilistic and marginal calibration. Finally, we use the trained model to generate PDFs for galaxies in our target datasets. 
We explore the performance of the full PDF estimations by examining both the point-estimates photometric redshifts and the calibration of their PDFs. For a given galaxy, the \phz value is defined as its respective PDF's most probable value (peak). Figure \ref{fig:dndz} presents the final photo-\textit{z} distribution, found with the full photo-\textit{z} PDF described in this appendix, in the 90$\%$ CI area of each gravitational wave event studied here. In order to highlight the overdensity regions, the uniform distribution in comoving volume $(dN/dz)_{\rm com}$ was subtracted from the photo-\textit{z} distribution $dN/dz$. 

To evaluate the quality of the point estimates, we employ the following metrics:

\begin{itemize}
    \item \textbf{Median Bias}: the bias, defined as $\Delta z = z_{\rm{phot}} - z_{\rm{spec}}$, directly measures the deviation of our estimations from the target values ($z_{\rm{spec}}$).

    \item \textbf{Scatter}: the Normalized Median Absolute Deviation $\sigma_{\rm{NMAD}}$, defined as
    
        \begin{equation}
            \sigma_{\rm{NMAD}} = 1.48\times {\rm{median}}\left(\left|\frac{\Delta z - {\rm{median}}(\Delta z)}{1+z_{\rm{spec}}}\right|\right),   
        \end{equation}

    is a standard measurement of the bias scattering \citep{Brammer_2008, Li_2022, Lima_2022}. We aim for $\sigma_{\rm{NMAD}}$ to be as low as possible.
    The choice of $\sigma_{\rm NMAD}$ instead of $\sigma_{68}$  is less sensitive to outliers. 

    \item \textbf{Outlier fraction}: outliers are defined as objects which
    
        \begin{equation}
            \frac{\Delta z}{1+z_{\rm{spec}}} > 0.15.   
        \end{equation}
            We define $\eta$ as being the fraction of outliers in any sub-sample of \phz estimations. This definition of an outlier follows the same approach as adopted in \cite{Ilbert_2006} and \cite{Lima_2022}.
        
\end{itemize}

    All the aforementioned metrics were computed using the objects with spectroscopic correspondence inside the 90\% confidence area of the events S231206cc, S230919bj, S230922g, GW191204\_171526, GW190924\_021846, GW200129\_065458, GW200202\_154313 and GW2003\\ 11\_115853. Additionally, with the exception of the odds constraints, the analysis refers only to the objects that satisfy the restrictions imposed to the events' \phz catalogues.

    \begin{figure*}
    \centering
      \centering
      \includegraphics[width=\linewidth]{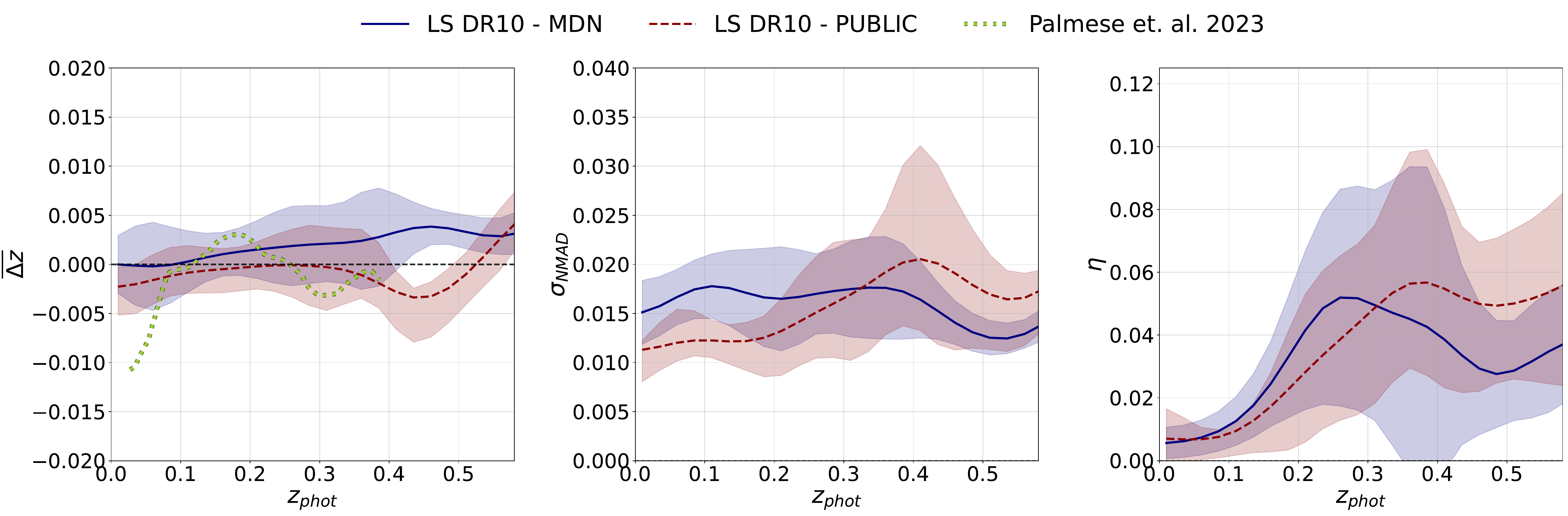}
      \caption {Mean values of the median bias ($\Delta z$), $\sigma_{\rm{NMAD}}$, and $\eta$ as a function of 0.025 width photo-\textit{z} bins. The solid lines depict the computed averages for all objects within each bin for each metric. The shadows represent the respective standard deviations.}
      \label{fig:metrics_z}
 \end{figure*}
    
   \begin{figure}
      \centering
      \includegraphics[width=\linewidth]{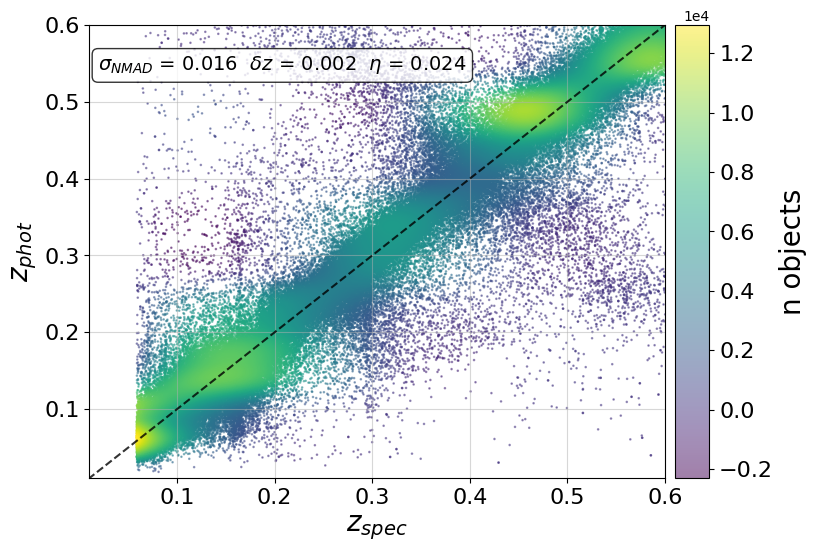}
      \caption{Photo-\textit{z}'s estimations versus the $z_{\rm{spec}}$ values. The colormap represents the density of objetcs, and on the top of the figure we displayed the point estimation metrics for the entire specz sample within the analysed GW and superevent regions.}
      \label{fig:xyplot}
    \end{figure}
    \noindent Figure \ref{fig:metrics_z} shows the median bias, scatter, and outlier fraction as a function of \phz bins with a width of 0.025. We generated a curve representing the mean value of these metrics within each \phz bin for each of the aforementioned GW events, using objects with available spec-z in each region. Subsequently, we averaged these curves across all events (solid lines) and computed the standard deviation between the values in each \phz bin (shaded regions).
    

Considering the measurements for both DELVE and LS surveys, we obtain $\sigma_{\rm{NMAD}}$ and an outlier fraction equal to 0.032 (0.016), 6.2(2.4)\% for DELVE (Legacy), respectively. These results show that our measurements using Legacy data outperforms those obtained using DELVE data. We use the Legacy public photometric redshifts in figure \ref{fig:metrics_z} catalogues for performance comparison in the point-like metrics. The LS public photometric redshifts were computed using the Random Forest method by \cite{zhou2020clustering}. In this method, they used r-band magnitude, \textit{g-r}, \textit{r-z}, \textit{z-W1}, and \textit{W1-W2} colours as input features.
We compare our results with the public \phz catalogue from LS-DR10 as well as the public redshifts from LS-DR9 \citep{Zou_2019} utilized in \cite{palmese2023}.
On the left panel of figure \ref{fig:metrics_z} the MDN method exhibits an improvement of the averaged median bias in the $0.01<z<0.1$ interval, compared both to the DR10 public available data and the data used in \cite{palmese2023}, which is relevant since a considerable amount of the objects in the event skymaps lies in that interval. Conversely, for higher redshifts, both curves are compatible within the scatter. Throughout the entire range of \phz, the bias for our method is very low and does not exceed $0.007$. It is also consistent with bias $0$ within the scatter until $z\sim 0.4$. However, it tends to be positive everywhere, indicating a weak cumulative overestimation effect. Additionally, in figure \ref{fig:metrics_z} the scatter and outlier fraction align with publicly available redshifts values, thereby affirming the reliability of our method in generating accurate \phz estimations. 

To validate the PDFs, we investigated the following metrics:

\begin{itemize}
    \item \textbf{PIT}: 
    
    the Probability Integral Transform \citep[PIT,][]{dawid_1984} represents the cumulative density function of the PDF up to the $z_{\rm{spec}}$ value for each galaxy. We can assess the quality of the PDFs by examining the PIT distribution for a representative sample of galaxies. The distribution of PIT values is expected to be uniform between 0 and 1 \citep{Mucesh_2021}, indicating that the $z_{\rm{spec}}$ values can be considered as random events generated from independent PDFs. A slope in the PIT distribution indicates a bias in the estimations, while the concavity of the distribution reveals whether the PDFs are over- or under-dispersed \citep{polsterer_2016}. 
    
    \item \textbf{Odds}:
    
    the odds \citep{benitez_2014jpas} measure the degree of confidence in the photo-$z$ estimate derived from a given PDF. It is defined as the probability of the redshift lying within an interval around the photo-$z$ value. For a given PDF, we can compute:
    
    \begin{equation}
        {\rm{Odds}} = \int^{z_{\rm{phot}}+0.06}_{{z_{\rm{phot}}-0.06}}\rm{PDF}(z)dz.
    \end{equation}

    The interval of $0.12$ around the \phz value was chosen based on the approach used by \citet{Coe_2006} in their analysis of photometric redshift estimates for galaxies in the Hubble Ultra Deep Field \citep{beckwith2006hubble}.
    The ideal distribution of odds for a galaxy sample well-represented by the training set should exhibit a pronounced peak near 1. However, it could also represent a distribution of under-dispersed PDFs. For this reason, we have to analyse the odds and PIT distribution simultaneously, in order to infer the quality of the PDFs.

    \item  \textbf{Coverage diagnostic}:  
    
    the coverage diagnostic (or High posterior density, HPD, diagnostic) stands as a well-established metric commonly employed to assess the quality of credible regions generated by simulation-based inference algorithms. Additionally, it provides a means to evaluate the accuracy of the estimated redshift distribution, as outlined in \cite{Dalmasso_2020}. 
    The fundamental concept involves evaluating the probability that a specified credible region within the inferred distribution contains the true value. This assessment provides insights into whether the estimated distribution is overconfident, calibrated, or underconfident \citep{hermans2022trust}. The coverage diagnostic was executed by selecting the pair (spec-\textit{z}, PDF) and sampling 1000 values from the photo-$z$'s PDF, thereby generating a frequency distribution with 1000 bins spanning the range from 0 to 1. 
    
    First, a credible region is defined for the estimated distribution using the highest density regions. This region represents the smallest area that contains at least $100(1 - \alpha)\%$ of the mass of the inferred photo-$z$ distribution, establishing an interval for a given credibility level ($1-\alpha$ ). The expected coverage is the frequency with which the true parameter (spec-\textit{z}) value falls within this highest density region, essentially indicating how often it falls inside the calculated interval.  
    If our model produces well calibrated distributions, we expect the spec-\textit{z} value should be contained inside the calculated interval from the $(1-\alpha)$ HPD region of the estimated distribution exactly $(1 - \alpha)\%$ of the time.
    If the coverage probability is less than the $(1 - \alpha)$ credibility level is the sign of under-estimation of the PDF's variance, and it could lead to unreliable approximations since it excludes physical values of \phz. Conversely, if the coverage probability is larger than the $(1 - \alpha)$ credibility level, then this indicates that the estimated PDFs are over-estimating their variance, in average.
  
\end{itemize}

\begin{figure*}
    \centering
    \begin{minipage}{.61\textwidth}
      \centering
      \includegraphics[width=1.\linewidth]{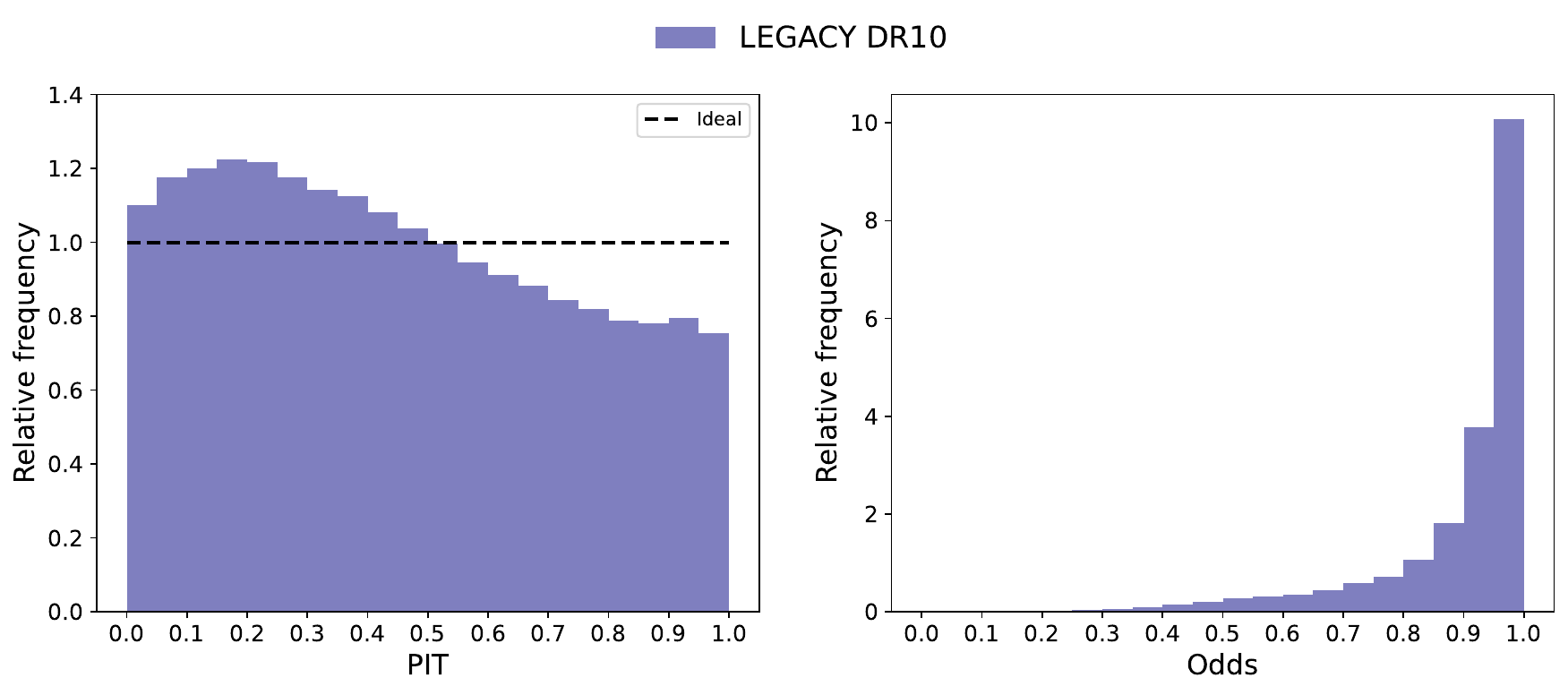}
      \caption{PIT and odds distributions for the objects with spectroscopic redshifts within the event regions. The odds distributions (right panel) remains for all objects (around 245k galaxies), meanwhile the PIT distribution (left panel) remains for the objects which obey $\rm{odds}>0.7$ beyond the errors constrains employed to the $H_0$ posterior estimations (around 130k galaxies).}
      \label{fig:pit_odds}
    \end{minipage}\hspace{2mm}
    \begin{minipage}{.36\textwidth}
      \centering
      \includegraphics[width=1.\linewidth]{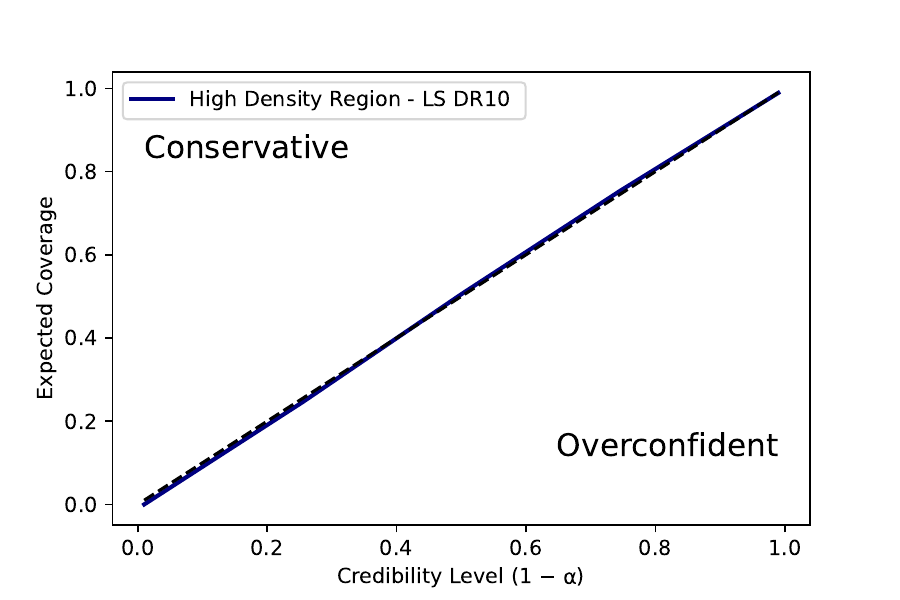}
      \caption{Coverage test results, where the blue solid line represents the resulting curve for our estimated PDFs and the black dashed line represents the ideal case of perfectly calibrated PDFs in terms of credibility.}
      \label{fig:coveragetest}
    \end{minipage}
\end{figure*}

The middle panel of figure \ref{fig:pit_odds} show the odds distribution for the same set of galaxies present on the point-estimate analysis, while the left panel illustrates the PIT distribution for a subset of these objects satisfying the constraint $\rm{odds}>0.7$. 
The latter exhibits a marginally concavity at lower values of PIT and a negative slope throughout higher values. Although not pronounced, this features collectively suggest a tendency towards positive bias, implying a slight overestimation of the most probable \phz's values. 
As discussed above, there is trending in the median bias curve (figure \ref{fig:metrics_z}) of overestimating \phz. This overestimation effect is further illustrated in figure \ref{fig:xyplot}, particularly evident around \phz equal to 0.48. 
Despite this anomaly at the higher redshifts, the scattering of \phz remains well-behaved, as expected. Consequently, the inclination of the PIT distribution might be attributed to a localized systematic bias rather than indicating any inherent global bias within the model. Although this could impact the analysis, it is worth stressing that our Bayesian analysis performs a bias correction based on this curve in the same manner as previous work by \cite{alfradique2024}. A similar shape of PIT distribution was also encountered by \cite{Isanto_2017} using MDN to estimate redshifts from images in the Sloan Digital Sky Survey Data Release 9.

The odds distribution shows a gradual increase in frequency with the odds values. This result suggests the absence of a systematically overconfident or overdispersed PDFs, indicating a high level of confidence in the estimated PDFs. From figure \ref{fig:coveragetest}, a perfectly calibrated estimated distribution aligns with a diagonal line (dashed black line), indicating an expected coverage probability matching the credibility level.  We notice that the expected coverage curve matches the diagonal, indicating that our model produces well calibrated \phz PDFs. This linear relation confirms the reliability of our methodology. 

In summary, our methodology estimates \phz with an accuracy comparable to those publicly available in LS-DR10, while also generating well-calibrated PDFs. It is noteworthy that our approach can predict redshifts in objects with detections in all \textit{g, r, i, z} bands, and in any combination of three bands containing detections in \textit{g} band.

\section{Method}
\label{sec:method}
Here, we outline the statistical methodology known as the dark siren approach, 
first idealized by \cite{schutz} and revisited in several works \citep{delpozzo, 2005ApJ...629...15H, macleod}. This methodology was initially detailed in \cite{chen17} and subsequently modified by \citep{darksiren1, palmese20_sts, palmese2023, alfradique2024}.
The method consists of using the Bayesian formalism to infer the Hubble constant parameter through the gravitational wave detection data $d_{\rm GW}$ and the $d_{\rm EM}$ set of photo-$z$ measurements of the possible host galaxies made using the deep learning algorithm presented in \ref{sec:a1_photoz}). As the GW and EM measurements are done independently, the joint GW and EM likelihood can be defined as the product of the two-individual likelihood, $p\left(d_{\rm GW}, d_{\rm EM}| H_{0}\right) = p\left(d_{\rm GW}|H_0\right) p\left(d_{\rm EM}|H_0\right)$. 

From the Bayesian framework, the $H_0$ posterior of one GW event can be written in the final form as

\begin{equation}
\begin{split}
p(H_0|d_{\rm GW}, d_{\rm EM}) &\propto \frac{p(H_0)}{\beta (H_0)} \sum_i \frac{1}{\mathcal{Z}_i} \int \de z_{i} ~\de \Delta z \,  \\
&\quad \times p(d_{\rm GW}|d_L(z_{i},H_0),\hat{\Omega}_i) p_i(d_{{\rm EM}, i}|z_{i},\Delta z)  \\
&\quad \times ~p(\Delta z) \frac{r^2(z_{i})\psi(z_{i})}{H(z_{i})} \, .
\end{split}
\label{eq:like3}
\end{equation}
where $p(H_0)$ is prior on $H_0$ which we assume to be flat over the range $[20,140]$ km/s/Mpc, $\beta (H_0)$ is the normalization factor that describes the selection effects in the measurement process, {$r(z)$ is the comoving distance}, $H\left(z\right)=H_{0}\left(\Omega_{m}\left(1+z\right)^3+1-\Omega_{m}\right)^{1/2}$ is the Hubble parameter in a Flat $\Lambda$CDM model, $\mathcal{Z}_i = \int p(d_{\rm EM} | z_i) r^2(z_i)/H(z_i) ~\de z_i$ are evidence terms that normalize the posterior where $p\left(d_{\rm EM} | z_i\right)$ is the EM likelihood marginalized over the photo-\textit{z} bias $\Delta z$, and $p\left(\Delta z\right)$ is the prior on the photo-$z$ bias that we measure from our photo-$z$ validation sample (see figure \ref{fig:metrics_z} in subsection \ref{sec:a1_photoz}). 
The last term on the equation above comes from the marginalization over the galaxies’ redshifts and sky positions, assuming that the galaxies are uniformly distributed in comoving volume with a merger rate evolution ,$\psi\left(z\right)$, following the Madau-Dickinson cosmic star formation rate \citep{Madau_2014}. In the following paragraphs, we will detail how the function $\beta\left(H_0\right)$ was computed. 

\begin{figure*}
    \centering
    \includegraphics[width=0.8\linewidth]{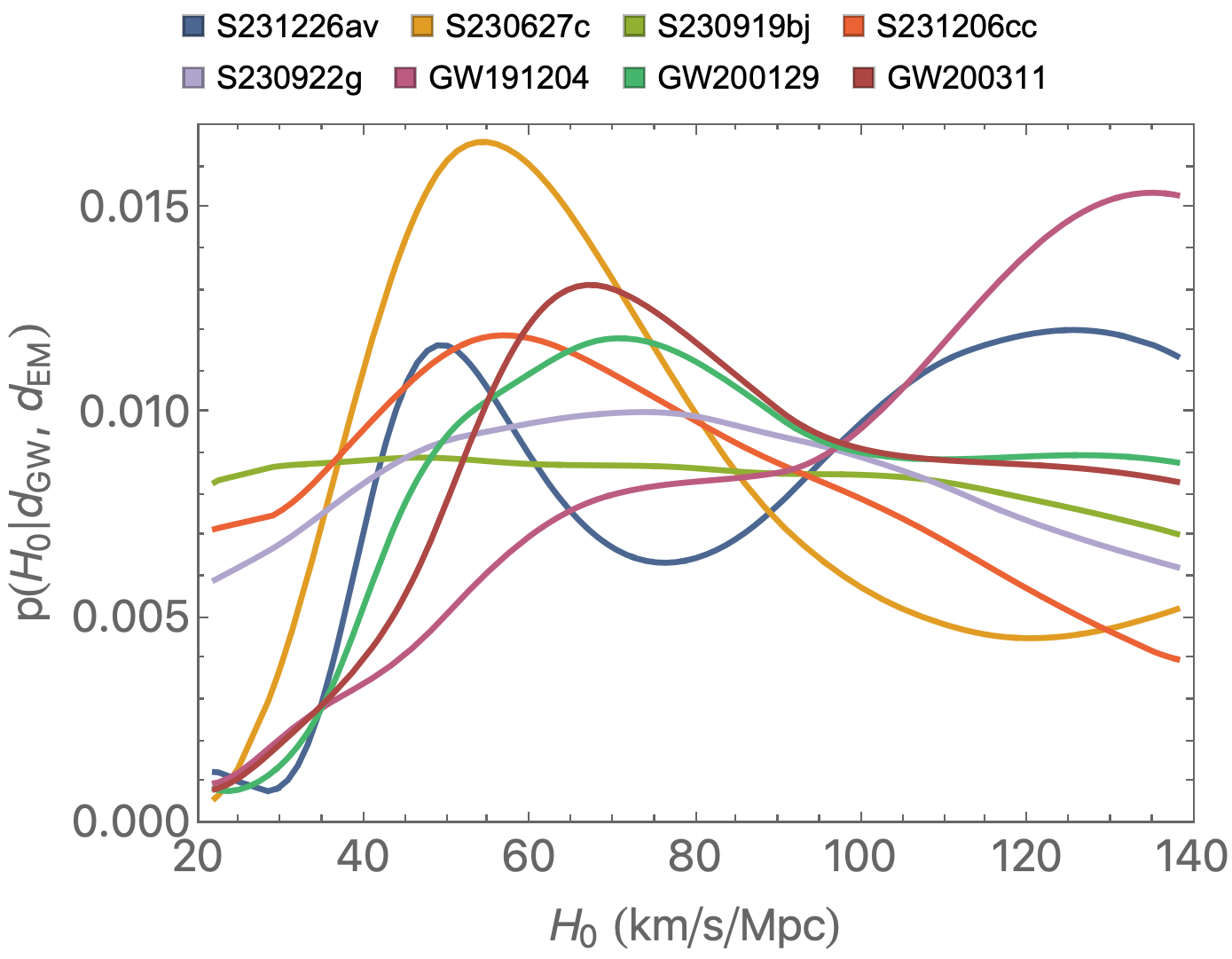}
    \caption{The Hubble constant posterior distributions for each dark siren considered in this work, which were found using the galaxies from Legacy survey. 
    }
    \label{fig:posteriors}
\end{figure*}

The GW and EM likelihoods are written as, respectively:
\begin{equation}
\begin{gathered}
p(d_{\rm GW}|d_{L}\left(z,H_0\right), \hat{\Omega}_{i})\propto \frac{p(\hat{\Omega}_{i})N(\hat{\Omega}_{i})}{\sqrt{2\pi}\sigma(\hat{\Omega}_{i})}\exp\left[-\frac{(d_{L}-\mu(\hat{\Omega}_{i}))^2}{2\sigma^{2}(\hat{\Omega}_{i})}\right], \\
p(d_{\rm EM}|z_{i}, \Delta z) = \prod_{i} p(\overline{z}_i|z_{i})p(\overline{z}_i|\Delta z_{i}),
\end{gathered}
\label{eq:likelihoods}
\end{equation}
where we explicitly consider the dependence of cosmological parameters on the luminosity distance, as measured by GW, and the solid angle $\hat{\Omega}_{i}$ corresponding to each observed galaxy \textit{i}. Following \cite{Singer_2016}, the GW likelihood is approximated by a Gaussian function. The EM likelihood is the product of the probability distribution function of the photometric redshift $\overline{z}_{k}$ for each \textit{k} galaxy, where we consider the correction of the shifted in redshift for the photo-$z$ biases, $\Delta z$, in the data. This phenomenon arises from the absence of a uniform distribution in redshift or colour beyond a certain magnitude or colour selection threshold, which causes the deep learning algorithm to oversample the redshift around the distribution peaks, thereby introducing systematic biases. The individual $H_0$ posterior distributions found in this work are presented (in colours) in figure \ref{fig:posteriors}.

The selection effects were computed based on the criteria outlined in \citep{chen17, gray2019cosmological}. This involves the joint gravitational wave-electromagnetic likelihood, which is marginalized over all conceivable GW and EM data. Assuming that the events are isotropically distributed on a large scale, this term can be written in its compact form as:
\begin{equation}
    \beta\left(H_0\right) = \int_{0}^{z_{\rm max}} p^{\rm GW}_{\rm sel}\left(d_{L}\left(z,H_{0}\right)\right)p\left(z\right)dz
\end{equation}

\noindent where $p_{\rm sel}^{\rm GW}$ is the probability of a GW event at a given luminosity distance $d_{L}$ to be detected, $z_{\rm max}$ is the maximum true redshift at which we can detect the host galaxies, and $p\left(z\right)$ is the galaxy catalogue distribution that for simplicity we assume to follow a uniformly distributed in comoving volume including the merger rate evolution described by the Madau-Dickinson star formation rate \citep{Madau_2014}. The function $p_{\rm sel}^{\rm GW}$ is equal to one if all events in a given redshift \textit{z} satisfy the detection condition, i.e. the detector network SNR is $>12$ and the localization volume satisfies our selection criteria ($A_{50\%}<1000\,\deg^2$ and $d_{L}<1500\,\rm{Mpc}$), and zero if none of the events located in a redshift \textit{z} satisfies those conditions. Therefore, the $p_{\rm sel}^{\rm GW}$ function is, in sum, an efficiency curve, i.e. a smooth function that falls from 1 to 0 over the redshift range for each $H_0$ value. It is important to mention that in the definition of $\beta \left(H_0 \right)$ the selection effects of EM data are incorporated assuming that all possible host galaxies are detected up to a maximum redshift, where for the Legacy catalog was adopted the value of $z_{\rm max}=0.6$.

The selection function was modelled in a similar way in the other previous dark sirens works \citep{alfradique2024,palmese2023,palmese20} by simulating 30,000 BBH mergers using \texttt{BAYESTAR} software \citep{bayestar,Singer_2016,Singer_supp} with the frequency domain approximant IMRPhenomD. We simulate the BBH mergers assuming that they follow a power law plus peak mass distribution with the same parameters as described in \cite{bom23bbh}. The spins distributions follow a uniform distribution between $(-1,1)$, and a uniform distribution in comoving volume assuming a dependency with the merger rate evolution described by the Madau-Dickinson star formation rate and fixing the Planck 2018 cosmology for twenty different $H_0$ values within our prior range. All the 30,000 injected events passed by the matched-filter analysis, where we computed the signal-noise-ratio (SNR) assuming the O4 sensibility curve as published by LIGO in Document P1200087\footnote{The data is available at \url{https://dcc.ligo.org/LIGO-T2000012/public}.} for the Hanford and Livingston LIGO network detectors. The network SNR defines the detection condition above 12 and at least 2 detectors have a single–detector SNR above 4. A Gaussian noise was added in all the measurements. Lastly, the \texttt{BAYESTAR} skymaps were reconstructed for each detection event, where we assume a luminosity distance that follows $\propto d_L^2$.
The last selection cut that we should consider is that events serving as dark sirens have at least 70\% of their 90\% CI comoving volume covered by the Legacy survey. As noted by \cite{palmese2023}, this selection cut can be ignored since the GW antenna pattern is not correlated with the survey sky footprint, which corresponds to selecting events isotropically. This is in agreement with the way that the selection function is computed, so we do not expect an strong $H_0$ dependence. Here, we also ignore this selection effect. However, it is worth noting that the value of $H_0$ changes the redshift range associated with a given luminosity distance horizon, consequently changing the SNR distribution of the source. Therefore, this may affect the size of the credible localization regions causing an impact on the validation of the survey's coverage condition. This effect will be investigated in more detail in a future work.

We analysed the effect that our choice of the power law plus peak mass distribution has on the selection function when compared to the result assuming a power law distribution. We found an average relative error of 7\%, which reflects a slight variation in the individual $H_{0}$ posteriors, where we observed a deviation of $1.2\,\rm{km/s/Mpc}$ at the peak of the $H_0$ posterior only for the event S230627c (the other events remained unchanged). We also examined the impact of our choice on the rate of BBH mergers by comparing our selection function with that found under the assumption that the merger rate density follows $(1+z)^{\kappa}$, with $\kappa$ fixed at the value inferred in \cite{gwtc3_population}. The results indicate a negligible relative error of $0.7\%$, implying a maximum variation of $0.9$ km/s/Mpc on the standard deviation of the individual posteriors. Therefore, the different choices of BBH population into the selection functions have a negligible effect in the combined $H_0$ constraint by standard dark sirens, lower than an order of magnitude of the present errors both in individual posteriors and the combined constraint.

We also accounted for a volume limited sample, we adopted a method similar to \citep{palmese20_sts,palmese2023,alfradique2024}. Initially, we set a maximum redshift of interest for each GW event based on its $90\%$ confidence interval in luminosity distance, averaged over the entire sky. This upper bound is translated into a maximum redshift using the highest $H_0$ value considered in our prior. For each event, we calculate the absolute magnitude that represents the galaxy sample's limiting apparent magnitude at that redshift. We then remove galaxies with absolute magnitudes lower than this threshold. Notably, while we employ a fiducial $\Lambda$CDM cosmology to determine these magnitudes, the $H_0$ dependence does not affect the results because the threshold value and the galaxies' absolute magnitudes scale with $H_0$ in the same way \citep{palmese20_sts}.



\section{Results and Discussion} \label{results}

\begin{figure*}
    \centering
    \includegraphics[width=0.8\linewidth]{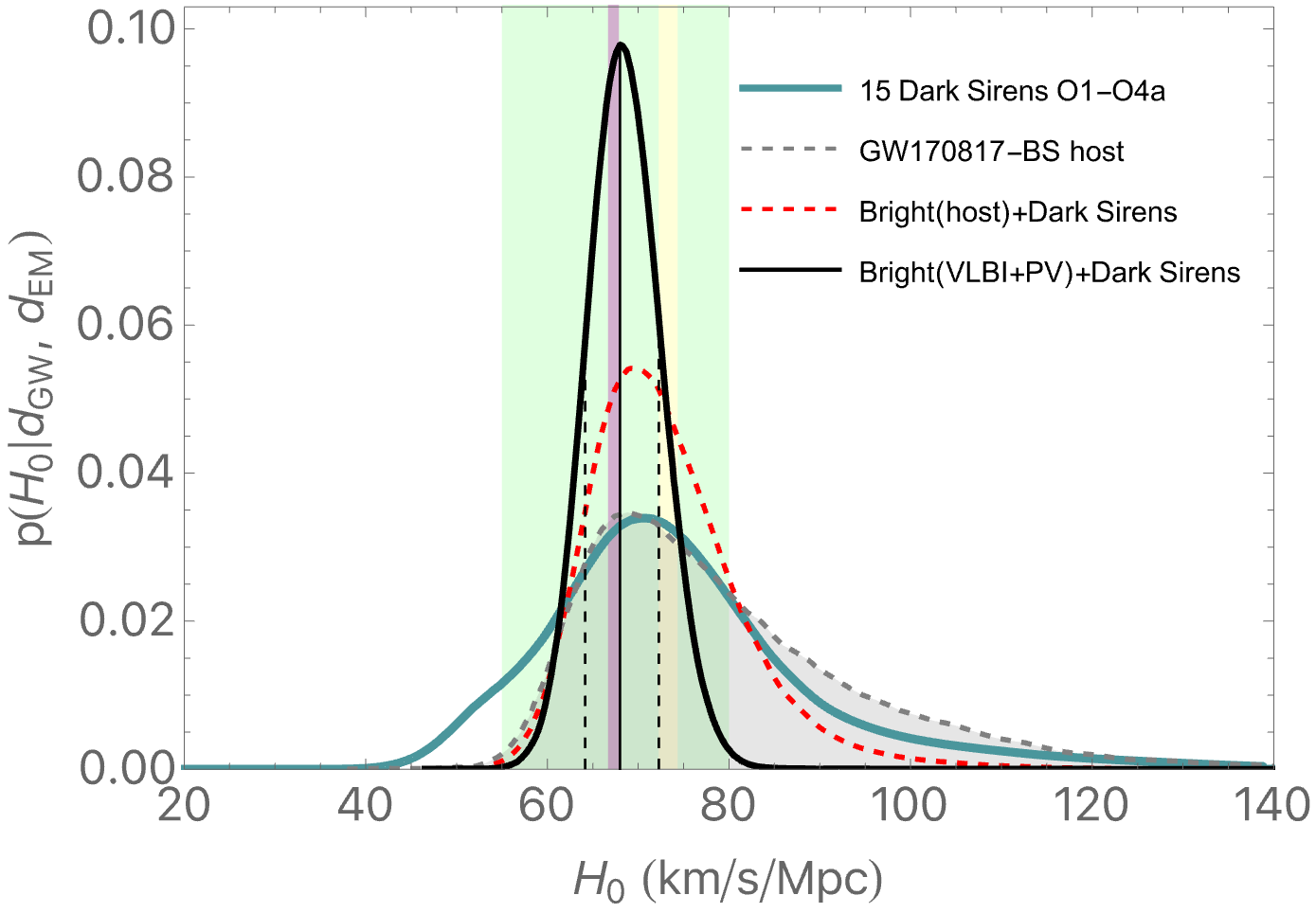}
    \caption{Hubble constant posterior distributions of the 15 dark sirens of O1-O4 observations. The dark green line shows the result from the combination of all the 15 dark sirens. The shaded gray posterior represents the GW170817 standard siren result, where only GW data is used, adapted from \citet{nicolaou2019impact}, which corrects the peculiar velocity to the constraint. The joint constraint from both the bright (i.e. GW170817, with the results from \citealt{nicolaou2019impact}) and the dark standard sirens is shown by the dotted red line. The black line is the 15 dark siren $H_0$ posterior (green line) combined with the results for the GW170817 bright siren found in \citet{Mukherjee_2021} using the GW+VLBI data with the peculiar velocity corrections, and the vertical dashed lines show the 68\% region for this posterior. For reference, we show the $1\sigma$ \citet{planck18} (shaded pink), \citet{riess2022} (R22, shaded light yellow), and the GWCT-3 (47 dark siren+ BBH population, \citealt{Abbott_2023}) constraints on $H_0$.}
    \label{fig:result}
\end{figure*}

In this section, we present the $H_0$ posterior produced using the dark siren methodology described in the previous section with the Legacy Survey photo-$z$’s PDFs constructed using the deep learning methodology from A24 and also described in section \ref{sec:a1_photoz}. We computed the $H_0$ posterior for five GW candidate BBH events detected in the LIGO O4a run and reanalysed three events from P23 with the updated skymaps, as they were not yet confirmed as gravitational wave events (see \S \ref{data}). The 68\% CI of the five O4a superevents represent about 69\% to 91\%, depending on the event analysed, of the 68\% CI of the prior width. The most constraining event from O4a is S230627c, because of its better localization, galaxy catalogue coverage, and quality of the photo-$z$ at the relevant redshift ranges. For the revisited gravitational wave events, despite the increase in 90$\%$ CI volume, all three events previously studied with the \texttt{bayestar} maps in P23 present an improvement of $\sim 18\%,~ 5\%$ and $9\%$ in 68\% CI, respectively.  
The main cause of these improvements is the quality of our photo-\textit{z} measurement (see figure \ref{fig:metrics_z}) as discussed in detail in the section \ref{sec:a1_photoz}. 

The majority of the $H_0$ posteriors show a clear peak, except for S231226av, which exhibit multiple peaks in different ranges of $H_0$ (raging from 49 to 136 km/s/Mpc). As already mentioned in previous works \citep{darksiren1, palmese20_sts, palmese2023, alfradique2024}, these peaks are associated with the presence of overdensities regions along the line of sight (see figure \ref{fig:dndz} in section \ref{sec:a1_photoz}). In the dark siren methodology, this is indicative of the host galaxy being more likely to be in these regions. Additionally, it is evident that certain events display a more prominent peak, which may stem from the presence of a prominent overdensity within an improved localization. In other words, the need for marginalization over fewer galaxies and/or the fact that the galaxies live at similar redshift, lead to a more informative $H_0$ posterior.

The $H_0$ posterior for GW191204  presents a considerable probability at $H_0\approx 130$ km/s/Mpc, which is associated with an overdensity of galaxies at $z\approx0.25$ (see figure \ref{fig:dndz} in section \ref{sec:a1_photoz}). We can also see that this $H_0$ posterior begins to decrease at the high–$H_0$ end, due to the absence of galaxies at high redshift (the galaxy density in figure \ref{fig:dndz} is negative at the high \textit{z} end, indicating the presence of an underdense region).

The combination of the 15 dark sirens from O1-O4a is depicted in figure \ref{fig:result} by the green line and detailed in table \ref{tab:results}. The maximum \textit{a posteriori} and the 68$\%$ CI is $H_{0} = 70.4^{+13.6}_{-11.7} \, \rm{km/s/Mpc}$.
As it can be seen, our results are consistent within 1$\sigma$ with Planck \citep{planck18} and the local Cepheid-supernova distance ladder \citep{riess2022}. This result agrees with the latest dark sirens study  using the BBH population and catalogue method \citep{Abbott_2023}, that found $H_0 = 67^{+13}_{-12}\, \rm{km/s/Mpc}$ for $47$ dark sirens from the third LIGO-Virgo-KAGRA GW transient catalogue \citep{gwtc3_published} and the \texttt{GLADE+} galaxy catalogue. A possible justification for the similar uncertainties, despite the difference in number of the events used, is the completeness and inhomogeneity of the \texttt{GLADE+} catalogue at the redshift of interest of the studied dark sirens. This contributes to the dependence between the final $H_0$ constrain and the BBH population assumptions observed in \cite{gwtc3_published}, and minimizes the galaxy method contribution to the final constraint. Furthermore, we also combine our dark siren results with the bright siren GW170817 from \cite{nicolaou2019impact}, which extended the analysis of the peculiar velocity effect in the $H_0$ measure presented in \cite{2017Natur.551...85A}. From this analysis, we find $H_{0} = 69.20^{+8.98}_{-5.85}\, \rm{km/s/Mpc}$. The dark siren information provides a notable reduction of 41$\%$ in the 68$\%$ CI and a 7$\%$ improvement in the relative precision. We further combine our final dark siren $H_0$ posterior with the bright siren analysis for GW170817 presented in \cite{Mukherjee_2021}, which corrects the peculiar velocity contribution in the measure of $H_0$ presented in \cite{Hotokezaka2019}. In that work, the superluminal motion measured by the Very Large Baseline Interferometry (VLBI, \citealt{Mooley2018}) and the afterglow data were used to measure the inclination angle of the GW170817 event. The addition of an independent EM observation helps break the degeneracy between the distance and the inclination angle, which is one of the main contributors to the uncertainty of the gravitational wave distance measurement. We find that the dark sirens information causes a reduction in the relative uncertainty from 18\% to 11\%, leading to our final constraint: $H_0=68.00^{+4.28}_{-3.85}\,\rm{km/s/Mpc}$. Furthermore, we combine our 15 dark sirens result with a more recent $H_0$ measurement \citep{Palmese2024}, which combines the GW170817 measurements with the electromagnetic counterpart associated with afterglow rather than the superluminal jet motion as done in \cite{Mukherjee_2021}. This result presents an agreement of 1.4$\sigma$ ($\sigma$ is uncertainty of $H_0$ from Planck data) with the $H_0$ measurements of Planck \citep{planck18}, indicating a reduction of 0.1$\sigma$ of the original result.  

To guarantee the use of the most appropriate galaxy catalogue in terms of coverage of the localization volumes of the gravitational wave events treated here and the quality of the estimated photo-$z$'s, we also analysed the measurements of $H_0$ for our dark sirens sample only using the DELVE catalogue. This allows us to evaluate if the $H_0$ measurement and photo-$z$ quality outperform our results obtained using the Legacy survey. 
The photometric redshifts for the DELVE galaxies were computed with the deep learning methods presented in A24, adapted for the redshift range of interest here. The results of the individual $H_0$ posterior distribution found with the Legacy and DELVE galaxy catalogs are in agreement; however, the combination of the 15 dark sirens found with DELVE leads to a higher $H_0$ peak value (approximately $2.0$ km/s/Mpc) and a lower uncertainty of $H_0$ ($\Delta\sigma_{H_0}\sim 0.5$ km/s/Mpc) compared to those presented in table \ref{tab:results}. This dark siren result combined with the bright siren analysis from \cite{Mukherjee_2021}, yields an $H_0$ measurement of $68.27^{+4.20}_{-3.91}\,\rm{km/s/Mpc}$, which is also in agreement with the Legacy Survey result. Although the DELVE results present a slightly more precise measurement than those found with the Legacy Survey, the quality of the metrics of the photo-$z$ measurements did not show the same robustness. The difference in reliability of photometric redshift measurements between DELVE and Legacy survey may be attributed to the absence of coadded data in DELVE, leading to lower SNR compared to the Legacy Survey catalogue. Furthermore, Legacy offers uniform sky coverage, enabling analysis of events across both the Southern and Northern skies. Given that the precision of photometric redshifts primarily influences $H_0$ measurements, we selected the Legacy results as our primary outcome.

We also investigated the effect of adding galaxy fakes to the galaxy catalogue for the S230919bj event. The presence of the galaxy fakes causes the peak of the posterior distribution to be slight shifted (a percent-level change) and the uncertainty is reduced by about 0.3$\%$. 
Concerning the combined result of the 15 dark sirens, the presence of the galaxy fakes leads to a $1\%$ increase in precision, and the result combined with GW170817 remains unchanged, which demonstrates the small impact that galaxy fakes have on the final conclusions of this work.

As other dark siren measurements, our $H_0$ measurement are valid under the Flat $\Lambda$CDM model and also presents a dependency on the background cosmology, since the events go beyond $z\sim 0.1$ where changes to $\Omega_m$ and other cosmological parameters have a more significant impact on $H_0$ estimates. Therefore, we check if our results change with the choice of the $\Omega_m$ value within the  $5\sigma$ interval found by the CMB measurement \citep{planck18}. There is a minor shift, with a relative difference of less than 4\% in the peak of the $H_0$ posterior distributions, and a relative difference less than 1\% in the $68\%$ CI. Therefore, the $H_0$ constraints presented here do not depend on the value of $\Omega_{m}$, as long as it agrees with the Planck constraints.

The $H_0$ posteriors presented in this work were computed considering the full redshift PDF, derived through the deep learning algorithm described \ref{sec:a1_photoz}, for each of the possible host galaxies. The application of a photo-\textit{z} PDF ensures the use of a more reliable galaxy distribution, first proposed in \cite{palmese20_sts}. This treatment is different from other previous dark sirens works~\citep{LVC_O2_StS, darksiren1, Abbott_2023}
that assumed a Gaussian approach for these distributions. 
In appendix \ref{app:gaussianapp}, we discuss the comparison between the results of these different methodologies; the $H_0$ posteriors' behaviour is slightly different, being more significant for closer events where the effect of marginalization over thousands of galaxies is minimized.
As already noted by \cite{palmese20_sts} and also seen here, the use of the Gaussian approximation causes the distribution of galaxies $dN/dz$ to be smoothed out, resulting in a flatter $H_0$ posterior. Despite these differences, the $H_0$ measurements are consistent with each other. 

\begin{table*}
\centering
\resizebox{\textwidth}{!}{
\begin{tabular}{cccc}
\hline\hline
Event(s) & Method(s)  & $H_0\,\left(\rm{km/s/Mpc}\right)$ & $\sigma_{H_0}\,\left(\rm{km/s/Mpc}\right)$  \\
\hline \hline
 O1-O3a $^1$ & Catalog & $67.3^{+ 27.6}_{-17.9}$ & $22.5$ $(34\%)$  \\
O1-O3 -- 8 dark sirens $^2$ & Catalog &  $79.8^{+19.1}_{-12.8}$ & $15.8$ $(20\%)$  \\
O1-O3 -- 47 dark sirens $^3$ & Catalog + BBH population & $67^{+13}_{-12}$ & $12.5$ $(18\%)$ \\
O1-O3 -- 47 dark sirens $^3$ & BBH population  & $67^{+14}_{-13}$ & $13.5$ $(20\%)$ \\
O1-O3 - 10 dark sirens $^4$ & Catalog  &$76.00^{+17.6}_{-13.4}$ &     $15.6$ $(20\%)$  \\
GW170817 $^5$ & Bright ($v_p$ corrected) & $68.80 ^{+ 17.3}_{-7.6}$ & $12.5(18\%)$  \\
GW170817 $^6$ & Bright (Chandra+HST+VLA) & $75.46^{+5.34}_{-5.39}$ & $5.36(7\%)$\\ 
GW170817 $^7$ & Bright ($v_p$+VLBI) & $68.3 ^{+4.6}_{-4.5}$ & $4.6(7\%)$ \\
\hline \hline
O1-O4a -- 15 dark sirens $^8$ & Catalog & $70.4^{+13.6}_{-11.7}$ & $12.6$ $(18\%)$ \\
O1-O4a -- 15 dark + 1 bright sirens $^8$ & Bright (host) + Catalog& $69.2^{+9.3}_{-5.6}$ & $7.5$ $(11\%) $  \\
O1-O4a -- 15 dark + 1 BS (EM) $^8$ & Bright (Chandra+HST+VLA) + Catalog& $74.3^{+5.1}_{-4.9}$ & $5.0$ $(7\%) $ \\ 
O1-O4a -- 15 dark + 1 BS (EM) $^8$ & Bright ($v_p$+VLBI) + Catalog& $68.0^{+4.4}_{-3.8}$ & $4.1$ $(6\%) $ \\ 
\hline \hline
\end{tabular}}
\caption{Hubble constant measurements using gravitational wave standard sirens from this work and previous works. $H_0$ values and uncertainties are given in km/s/Mpc, and $H_0$ priors are flat, unless otherwise stated. The uncertainty from the flat prior only is derived by assuming the same $H_0$ maximum found in the analysis. Quoted uncertainties represent 68\% HDI around the maximum of the posterior. The ``$\sigma_{H_0}/\sigma_{\rm prior}$'' column shows the 68\% CI from the posterior divided by 68\% CI of the
prior width. (1) \citet{finke2021cosmology}, (2) \citet{palmese2023}, (3) \citet{Abbott_2023}, (4) \citet{alfradique2024}, (5) adapted from \citet{nicolaou2019impact}, (6) \citet{Palmese2024}, (7) \citet{Mukherjee_2021}, and (8) this work.} 
\label{tab:results}
\end{table*}

\section{Conclusions}\label{conclusions}

In this contribution, we use the data from the current best-localized and covered GW events from LVK O4a observing run to derive a dark siren measurement of the Hubble constant using the galaxy catalogue method and precise photometric redshifts. We obtained $H_0=70.4^{+13.6}_{-11.7}$ km/s/Mpc, i.e. a $\sim 18\%$ uncertainty on $H_0$ from dark sirens based on the catalogue method alone and a total of $15$ sirens. This is, at the best of our knowledge, an unprecedented precision for the catalogue method and for the dark siren approach. 

We combine our results from $15$ dark sirens with recent constraints over the one bright standard siren available GW170817, considering the constraints to the viewing angle from VLBI and the host galaxy peculiar velocity \citep{Mukherjee_2021}. We obtained $H_0=68.00^{+4.28}_{-3.85}$ km/s/Mpc, representing a $6\%$ measurement of $H_0$, which reduces the previous constraint uncertainty by approximately 10\%. We note that the precision and uncertainty of the $15$ dark sirens are similar to the GW170817 constraint without the viewing angle constraints from electromagnetic observations. It is worth noting that the current results are derived under the assumption of a Flat $\Lambda$CDM scenario.

Our current results emphasize that a combination of well-localized dark sirens and high-quality photometric redshifts can achieve a competitive $H_0$ constraint from gravitational waves. In particular, considering the absence of high confidence binary neutron star merger detections during O4a, the BNS merger rate can be in the lower-end of previous estimations. Therefore, the number of well-localized BBHs detections could be one order of magnitude higher than that of BNS. Furthermore, neutron star-black hole (NSBH) systems do not present substantial promise as multimessenger sources \citep{nsbh}. Nonetheless, it is prudent to acknowledge that these detection rates may change significantly in the near future, and the emergence of a singular, observable BNS event with an electromagnetic counterpart could potentially offer more compelling constraints than a dozen dark standard sirens. 

The current constraints from dark sirens by the catalogue method only, achieve a precision of $\sim 18\%$ and, and in combination with bright sirens and additional constraints on the viewing angle, can achieve $\sim 6\%$. As the number of dark sirens events increase and we get closer to the level of statistical precision required to arbitrate the Hubble tension of $\sim 2\%$ detailed studies to address potential systematics not included in this work should be carried out, especially considering different formation channels for BBH populations and catalogue depth \citep{gray2019cosmological, Mastrogiovanni2023}.   


 

\section*{Acknowledgements}

\noindent CRB acknowledges the financial support from CNPq (316072/2021-4) and from FAPERJ (grants 201.456/2022 and 210.330/2022) and the FINEP contract 01.22.0505.00 (ref. 1891/22). This material is based upon work supported by NSF Grant No. 2308193. AP thanks Constantina Nicolaou for sharing the GW170817 posterior. The authors made use of Sci-Mind servers machines developed by the CBPF AI LAB team and would like to thank A. Santos, P. Russano, and M. Portes de Albuquerque for all the support in infrastructure matters. 

\noindent The Legacy Surveys consist of three individual and complementary projects: the Dark Energy Camera Legacy Survey (DECaLS; NSF's OIR Lab Proposal ID 2014B-0404; PIs: David Schlegel and Arjun Dey), the Beijing-Arizona Sky Survey (BASS; NSF's OIR Lab Proposal ID 2015A-0801; PIs: Zhou Xu and Xiaohui Fan), and the Mayall z-band Legacy Survey (MzLS; NSF's OIR Lab Proposal ID 2016A-0453; PI: Arjun Dey). DECaLS, BASS and MzLS together include data obtained, respectively, at the Blanco telescope, Cerro Tololo Inter-American Observatory, The NSF's National Optical-Infrared Astronomy Research Laboratory (NSF's OIR Lab); the Bok telescope, Steward Observatory, University of Arizona; and the Mayall telescope, Kitt Peak National Observatory, NSF's OIR Lab. The Legacy Surveys project is honored to be permitted to conduct astronomical research on Iolkam Du'ag (Kitt Peak), a mountain with particular significance to the Tohono O'odham Nation.

\noindent The NSF's OIR Lab is operated by the Association of Universities for Research in Astronomy (AURA) under a cooperative agreement with the National Science Foundation.

\noindent This project used data obtained with the Dark Energy Camera (DECam), which was constructed by the Dark Energy Survey (DES) collaboration. Funding for the DES Projects has been provided by the U.S. Department of Energy, the U.S. National Science Foundation, the Ministry of Science and Education of Spain, the Science and Technology Facilities Council of the United Kingdom, the Higher Education Funding Council for England, the National Center for Supercomputing Applications at the University of Illinois at Urbana-Champaign, the Kavli Institute of Cosmological Physics at the University of Chicago, Center for Cosmology and Astro-Particle Physics at the Ohio State University, the Mitchell Institute for Fundamental Physics and Astronomy at Texas A\&M University, Financiadora de Estudos e Projetos, Fundacao Carlos Chagas Filho de Amparo, Financiadora de Estudos e Projetos, Fundacao Carlos Chagas Filho de Amparo a Pesquisa do Estado do Rio de Janeiro, Conselho Nacional de Desenvolvimento Cientifico e Tecnologico and the Ministerio da Ciencia, Tecnologia e Inovacao, the Deutsche Forschungsgemeinschaft and the Collaborating Institutions in the Dark Energy Survey. The Collaborating Institutions are Argonne National Laboratory, the University of California at Santa Cruz, the University of Cambridge, Centro de Investigaciones Energeticas, Medioambientales y Tecnologicas-Madrid, the University of Chicago, University College London, the DES-Brazil Consortium, the University of Edinburgh, the Eidgenossische Technische Hochschule (ETH) Zurich, Fermi National Accelerator Laboratory, the University of Illinois at Urbana-Champaign, the Institut de Ciencies de l'Espai (IEEC/CSIC), the Institut de Fisica d'Altes Energies, Lawrence Berkeley National Laboratory, the Ludwig-Maximilians Universitat Munchen and the associated Excellence Cluster Universe, the University of Michigan, the National Optical Astronomy Observatory, the University of Nottingham, the Ohio State University, the University of Pennsylvania, the University of Portsmouth, SLAC National Accelerator Laboratory, Stanford University, the University of Sussex, and Texas A\&M University.

\noindent BASS is a key project of the Telescope Access Program (TAP), which has been funded by the National Astronomical Observatories of China, the Chinese Academy of Sciences (the Strategic Priority Research Program "The Emergence of Cosmological Structures" Grant \# XDB09000000), and the Special Fund for Astronomy from the Ministry of Finance. The BASS is also supported by the External Cooperation Program of Chinese Academy of Sciences (Grant \# 114A11KYSB20160057), and Chinese National Natural Science Foundation (Grant \# 11433005).

\noindent The Legacy Survey team makes use of data products from the Near-Earth Object Wide-field Infrared Survey Explorer (NEOWISE), which is a project of the Jet Propulsion Laboratory/California Institute of Technology. NEOWISE is funded by the National Aeronautics and Space Administration.

\noindent The Legacy Surveys imaging of the DESI footprint is supported by the Director, Office of Science, Office of High Energy Physics of the U.S. Department of Energy under Contract No. DE-AC02-05CH1123, by the National Energy Research Scientific Computing Center, a DOE Office of Science User Facility under the same contract; and by the U.S. National Science Foundation, Division of Astronomical Sciences under Contract No. AST-0950945 to NOAO.

\noindent The Photometric Redshifts for the Legacy Surveys (PRLS) catalog used in this paper was produced thanks to funding from the U.S. Department of Energy Office of Science, Office of High Energy Physics via grant DE-SC0007914.

\section*{Data Availability}
 
The data underlying this article will be shared on reasonable request to the corresponding author.



\bibliographystyle{mnras}
\bibliography{references} 

\appendix
\section{Comparison with $H_{0}$ constraints from point estimates of photometric redshifts}\label{app:gaussianapp}

\begin{figure*}
    \centering
    \begin{subfigure}{}
        \includegraphics[width=2in]{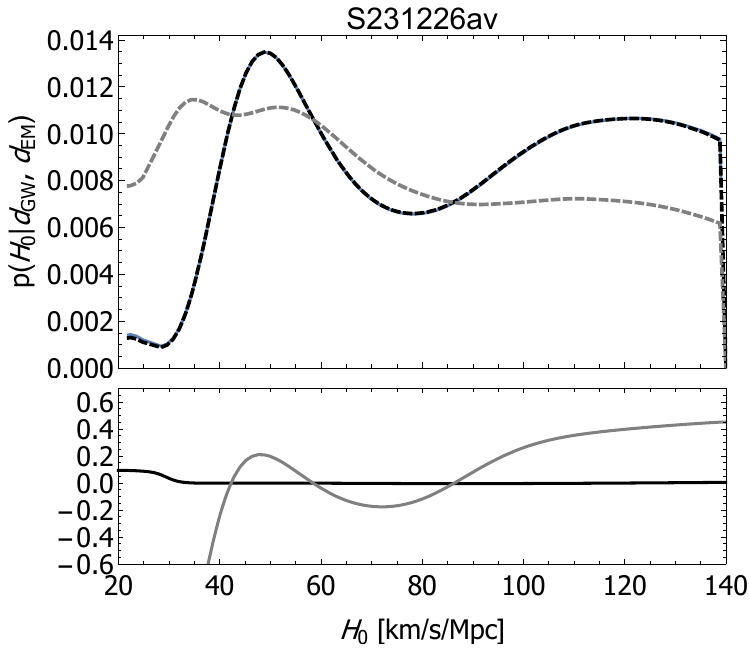}
    \end{subfigure}
    \begin{subfigure}{}
        \includegraphics[width=2in]{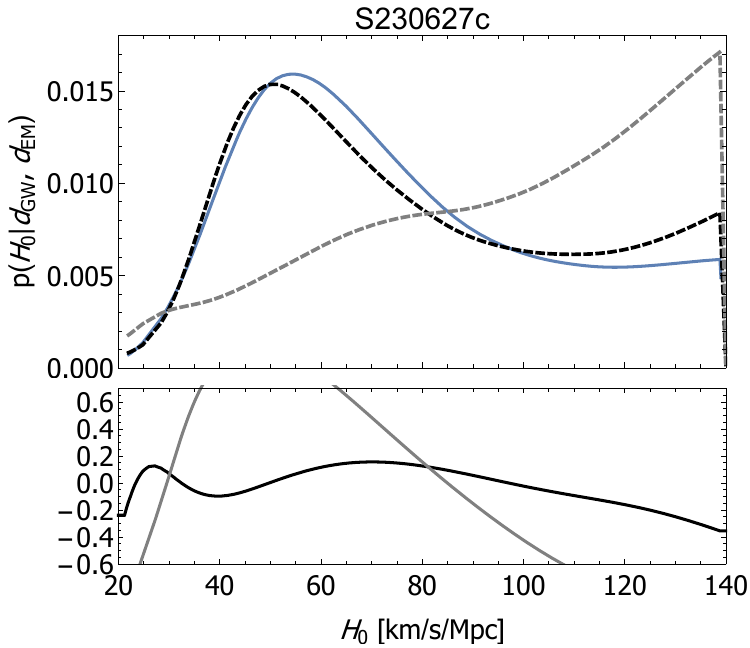}
    \end{subfigure}
    \begin{subfigure}{}
        \includegraphics[width=2in]{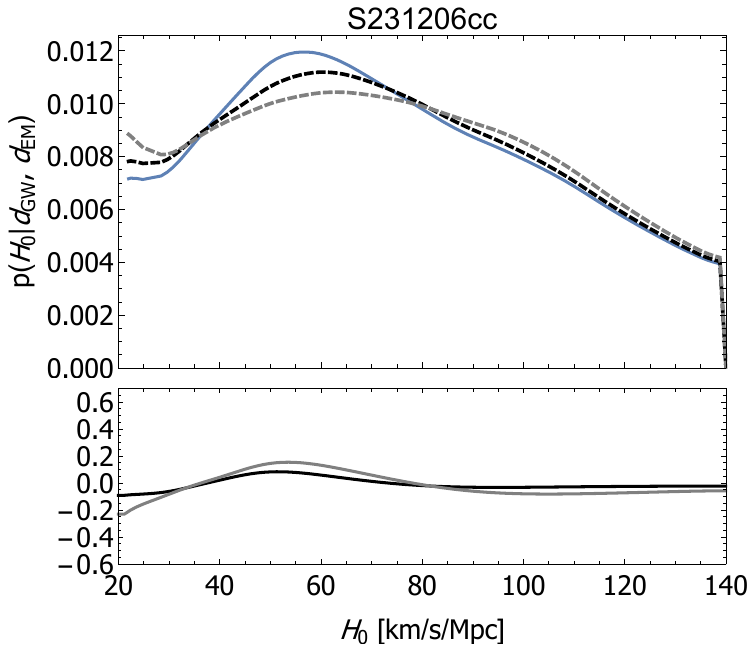}
    \end{subfigure}

    \begin{subfigure}{}
        \includegraphics[width=2in]{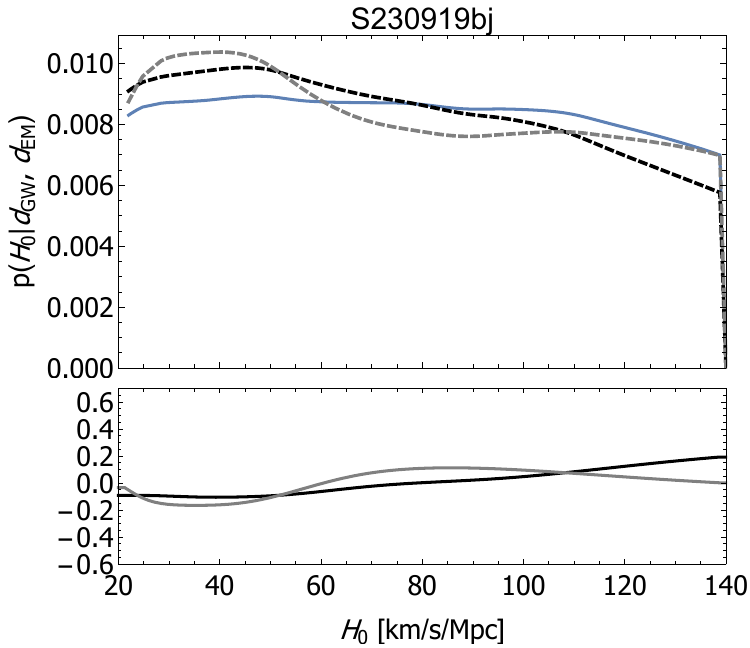}
    \end{subfigure}
    \begin{subfigure}{}
        \includegraphics[width=2in]{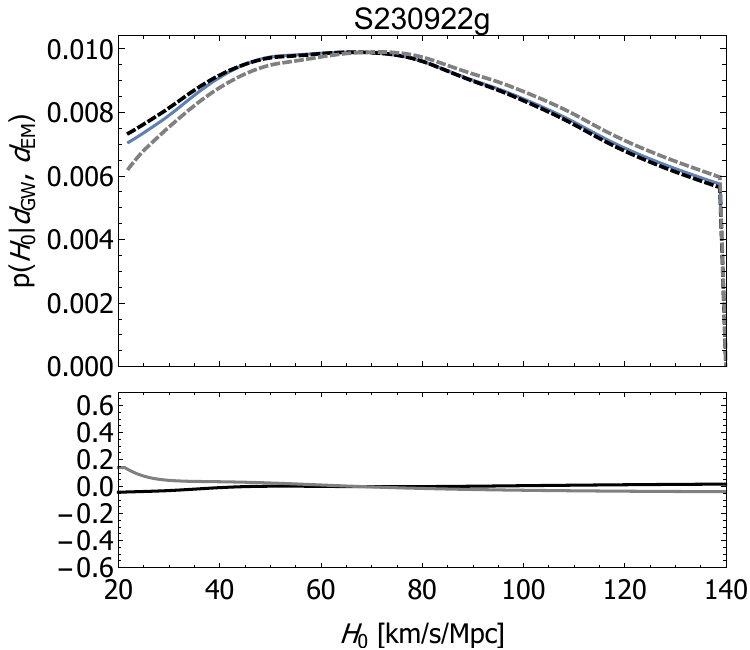}
    \end{subfigure}
    \begin{subfigure}{}
        \includegraphics[width=2in]{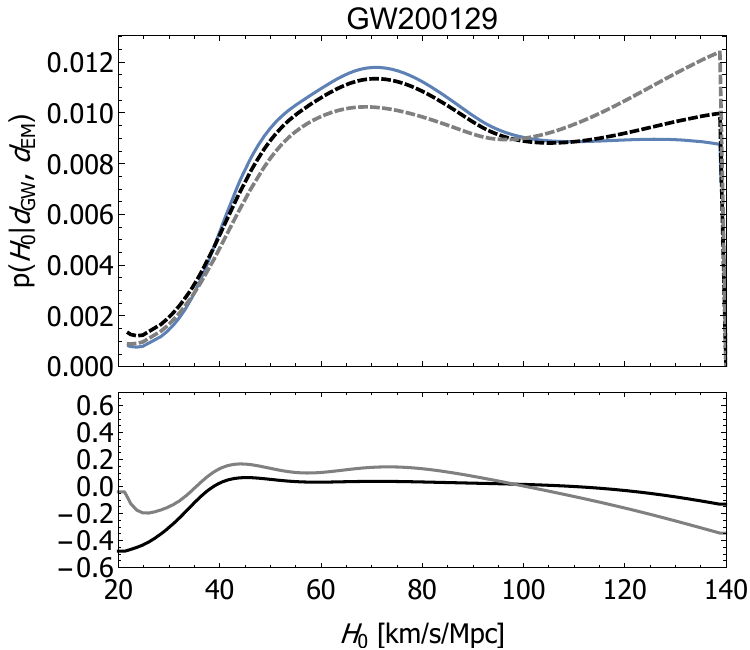}
    \end{subfigure}

    \begin{subfigure}{}
        \includegraphics[width=2in]{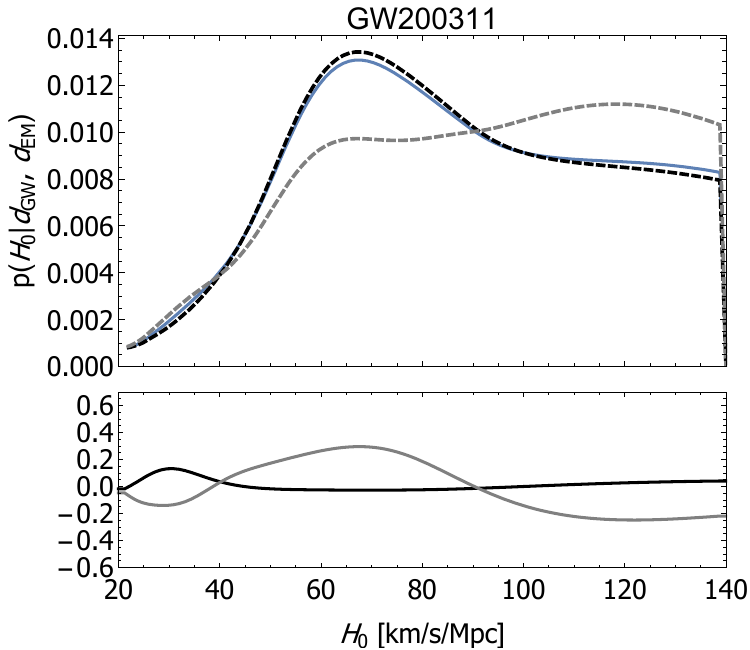}
    \end{subfigure}
    \begin{subfigure}{}
        \includegraphics[width=2in]{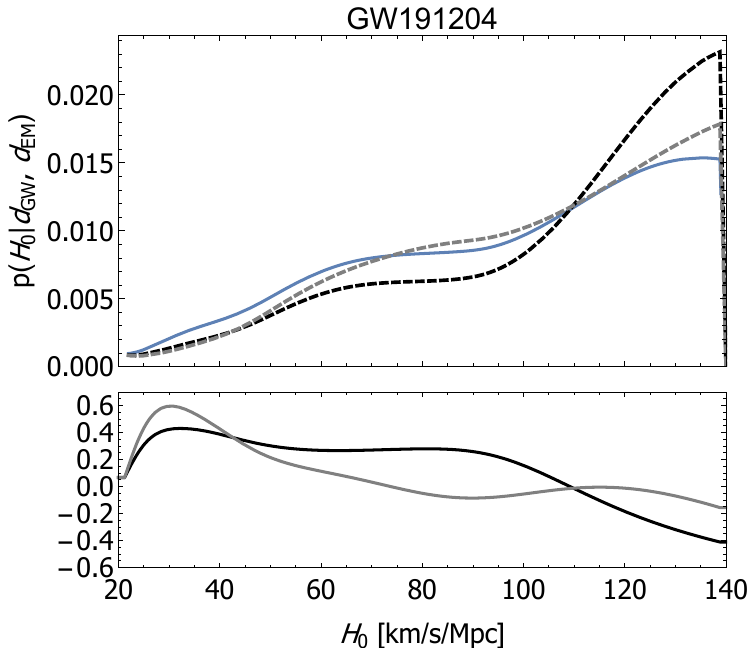}
    \end{subfigure}
    
    \caption{Comparison between the Hubble constant posterior distributions found using the full galaxies redshift PDFs (blue solid line) and a Gaussian approximation (dashed line: (1) black colour using the means and standard deviation of the full PDFs, and (2) gray colour using the photo-$z$ measurements made by the public Legacy DR10.1) for all the dark sirens studied here. The bottom panels show the residuals value between the two curves shown on the top panel, calculated as $2\times( \rm{solid\, line}- \rm{dashed\, line})/( \rm{solid\, line}+ \rm{dashed\, line})$.}
    \label{fig:gaussianeff}
\end{figure*}

During this work, the $H_0$ posterior calculation was performed using the full photo-$z$ PDFs estimated through the deep learning method described in section \ref{sec:a1_photoz}. In this appendix, we will analyze the effect on the $H_0$ posterior originated by the choice of the galaxies photo-$z$ PDFs. We will consider the photo-$z$ PDF estimated by the MDN technique described in section \ref{sec:a1_photoz}, which we will call the full photo-$z$ PDF, or a Gaussian approximation, where we approximate each full photo-$z$ PDFs to a Gaussian function whose mean coincides with the peak value and the standard deviation is equal to that found with the full photo-$z$ PDF.
In this context, the Gaussian approximations are constructed with the same peak and standard deviation as those found with the PDFs estimated with the MDN. We include the results of the public Legacy \footnote{The photo-$z$ were computed using the Random Forest technique with data from the Legacy DR10.1, these results are available in \url{https://www.legacysurvey.org/dr10}.} in the figures. The top panel of figure \ref{fig:gaussianeff} shows the posterior of $H_0$ found considering the full photo-$z$ PDF (blue solid curve) and the Gaussian approximation (black dashed curve for the full PDF, this work, and the Legacy DR10.1 public results in gray dashed curve), along with the residuals curves (see the bottom panels). 

Comparing the results of the $H_0$ posterior generated using the Gaussian approach (black dashed curve) and the full photo-$z$ PDF, we note that the results are in agreement at the percentage level for the more distant events, where the effect is suppressed by marginalization over a larger number of galaxies due to their larger localization volume. The impact of using the Gaussian approximation is more evident for the events S230627c, GW200129 and GW191204, which present a discrepancies of approximately 4\%, 16\%, and 40\%, respectively, in the peak distribution and can reach values $>13\%$ at the ends. As expected, the Gaussian approximation makes the peak of the $H_0$ posterior wider, implying a less precise $H_0$ measurement. This result is a consequence of the smoothing of overdensity regions in the photo-$z$ distribution. Although the results do not indicate that the choice of photo-$z$ PDF leads to disagreement in $H_0$ measurements, future dark siren measurements may reach a level of precision where the differences between these results could be statistically significant. We can also see that the Legacy public results leads to, in most events, a less restrictive $H_0$ posterior compared to those achieved through the full photo-$z$ PDFs, resulting in an increase of up to 6\% in the $H_0$ uncertainty.

As shown by \cite{darksiren1}, the choice for a wider redshift cut causes galaxies to be added at deeper redshift, whose photo-$z$ PDF has significant Gaussian tails at high redshift, which implies an increase in the $H_0$ posterior at high $H_0$. However, the full photo-$z$ PDF is able to reduce (the residual values, between the posterior considered a LIGO/Virgo luminosity distance posterior of 90\% and 99.7\%, at the high-$H_0$ end reduced by $\approx 25\%$ compared to the result achieved with the Gaussian approximation) this dependence of the $H_0$ posterior behavior with redshift cut, which evidences its advantage against the Gaussian approximation in performing a measurement of $H_0$ free of any systematic imposed by assumptions made in the methodology.  

\begin{figure*}
      \includegraphics[width=\linewidth]{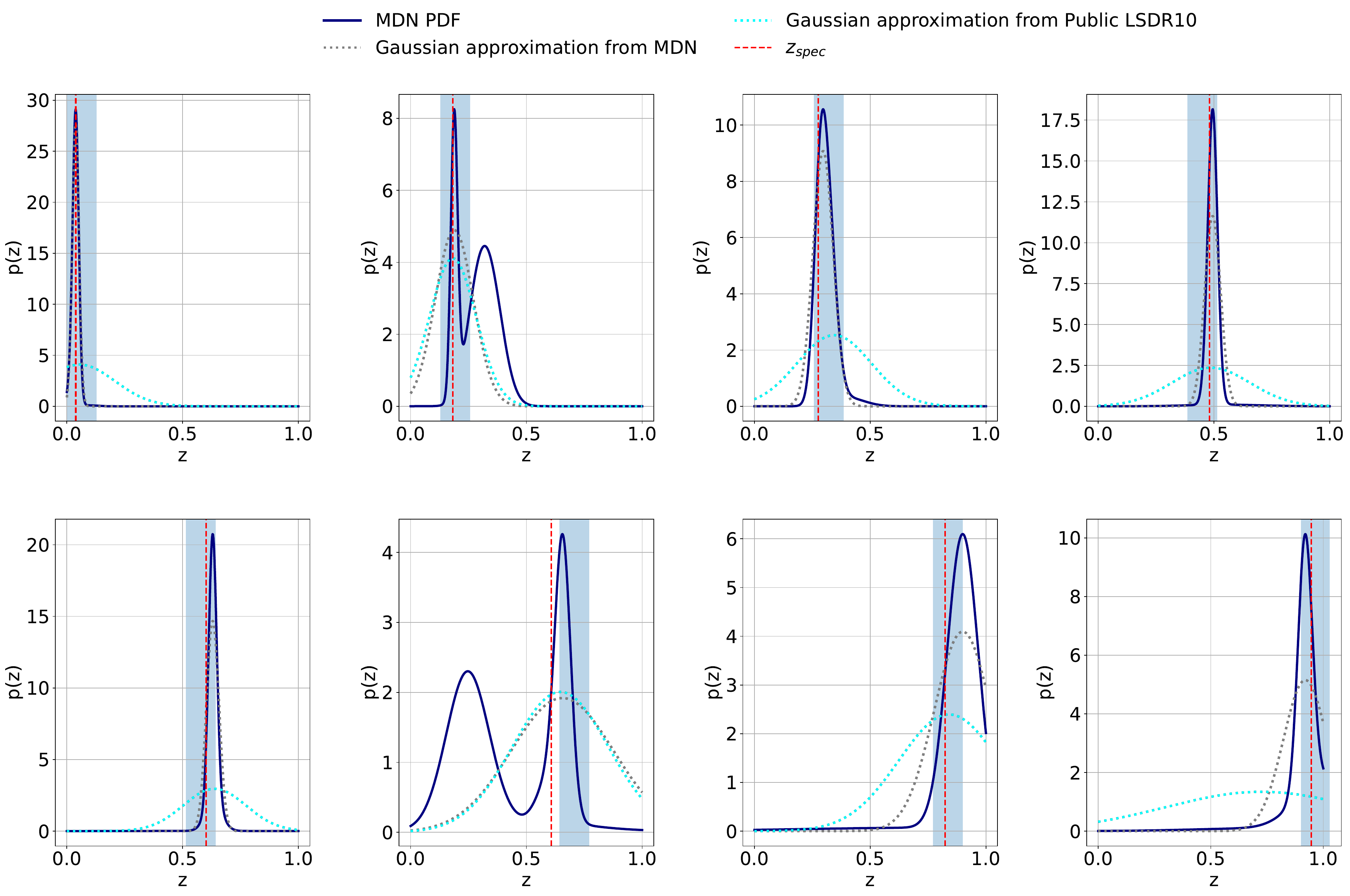}
      \caption{Photometric redshift PDFs estimated with the MDN technique (dark blue line), Gaussian approximation of MDN photo-$z$ PDF (dashed curve in cyan), and Gaussian approximation using the photo-$z$ estimated by the Public LSDR10 (dashed curve in gray) for different redshifts (increases from left to right). The vertical dashed red line represents the corresponding $z_{\rm spec}$ value.}
      \label{pdfsfull}
 \end{figure*}

In figure \ref{pdfsfull}, we present the photo-$z$ PDFs estimated by the deep learning method (dark blue curve) and the Gaussian approximation (dashed curve in light blue) for galaxies at different redshifts. We notice that the full photo-$z$ PDF has a tendency to be narrower than the Gaussian approximation, which favours a more precise measurement of $H_0$. Additionally, we see that the difference between the estimated PDF and the Gaussian approximation becomes more evident at higher redshifts. This highlights the importance of the MDN technique, especially for galaxies located at deep redshifts, which are already used for the LIGO/Virgo GW dark sirens and will be increasingly indispensable in the coming years where it is expected to observe increasingly distant GW events. 

\begin{figure}
     \centering
     \includegraphics[width=1\linewidth]{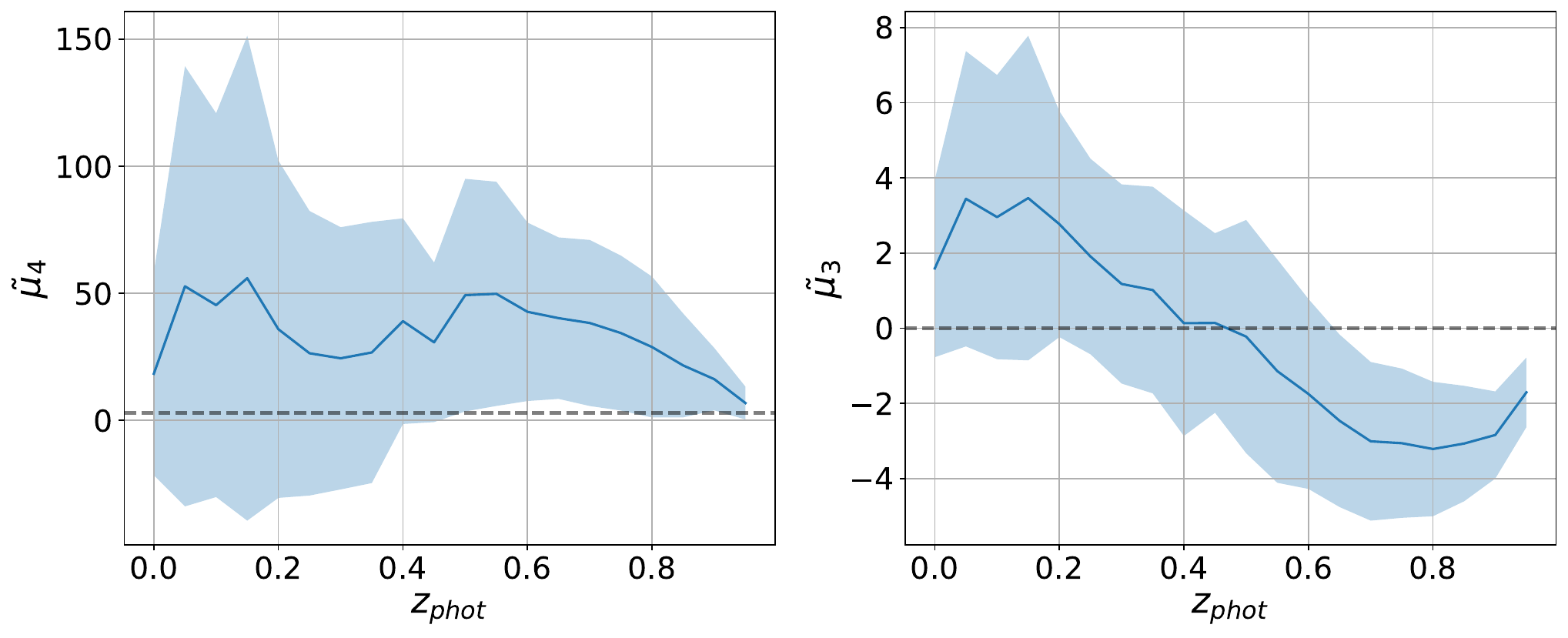}
     \caption{Kurtosis (left panel) and skewness (right panel) values of the photo-$z$ PDFs, estimated with the MDN technique, as a function of $z_{\rm phot}$. The solid line represents the mean value of each quantity, and the shaded region represents the respective standard deviations computed in photo-$z$ bins of size 0.05.}
     \label{kurtosis}
 \end{figure}

We can also observe that the full photo-$z$ PDFs become increasingly non-Gaussian at deeper redshifts. To quantify this non-Gaussianity, in figure \ref{kurtosis} we present the relation between the kurtosis (fourth standardized moment $\Tilde{\mu}_{4}$, that is presented in the left panel) and the skewness (third standardized moment $\Tilde{\mu}_{3}$ shown in the right panel) of the full PDFs with $z_{\rm phot}$. We compute the mean (solid line) and the standard deviation (shaded region) of this quantities in photo-$z$ bins of size 0.05. The values of kurtosis and skewness are distinct from those of a Gaussian distribution (kurtosis and skewness equals to 3 and 0, respectively) across almost the entire $z_{\rm phot}$ interval. We note a tendency of kurtosis values to decrease with increasing $z_{\rm phot}$, indicating the broadening of PDFs, which implies less precise $z_{\rm phot}$ measurements. This behaviour reflects the expected difficulty of obtaining precise measurements of distant galaxies. The skewness results indicate that the full PDFs are right-skewed at low $z_{\rm phot}$, decreasing with increasing $z_{\rm phot}$ until reaching a value of zero (indicating a symmetric PDF) at $z_{\rm phot}\approx0.5$, after which they begin to be left-skewed.

\bsp	
\label{lastpage}
\end{document}